\documentclass[10pt,letterpaper]{article}
\usepackage{opex3}
\usepackage{subfigure}
\usepackage{cite}

\begin{document}

\title{Imaging velocities of a vibrating object by stroboscopic sideband holography}

\author{F.~Verpillat,\textsuperscript{1} F.~Joud,\textsuperscript{1} M.~Atlan,\textsuperscript{2}  and M.~Gross\textsuperscript{3*}}

\address{\textsuperscript{1} Laboratoire Kastler Brossel: UMR 8552 CNRS- ENS- UPMC, 24 rue Lhomond 75231 Paris 05, France\\
\textsuperscript{2} Institut Langevin, UMR 7587 CNRS ESPCI ParisTech,
1, rue Jussieu, 75005 Paris\\
\textsuperscript{3} Laboratoire Charles Coulomb:  UMR 5221 CNRS, Universit\'{e} Montpellier 2, 34095 Montpellier, France}

\email{michel.gross@univ-montp2.fr\textsuperscript{*}}
%
%

\begin{abstract}
We propose here to combine sideband holography with stroboscopic illumination synchronized
with the vibration of an object.  By sweeping the  optical frequency of the reference beam  such a way the holographic detection is tuned on
the successive sideband harmonic ranks, we are able to image the instantaneous velocities of the object. Since the stroboscopic illumination is made with an electronic  device, the method is compatible with fast (up to several MHz) vibration motions. The method is demonstrated  with a vibrating clarinet reed excited sinusoidally at 2 kHz, and a stroboscopic illumination with cyclic ratio   $0.15$.  Harmonic rank up to $n=\pm 100$ are detected, and  a movie of the instantaneous velocities  is reported.
\end{abstract}


\bibliographystyle{unsrt}

\section{Introduction}

There is a big demand for full field vibration measurements, in particular in industry. Different holographic techniques are able to image and analyse such vibrations.
Double  pulse holography \cite{pedrini1995digital,pedrini1997digital} records a double-exposure hologram with time separation in the 1...1000 $\mu$s range, and  measures the instantaneous velocity of a vibrating object  from the phase difference. The method  requires a quite costly double pulse ruby laser system, whose repetition rate is low. Multi pulse holography \cite{pedrini1998transient} is able to analyse transient motions, but the setup is still heavier (4 pulses laser, three cameras).

The development of fast CMOS camera makes possible to analyse vibration efficiently by triggering the camera on the motion in order to record a sequence of holograms that allows to track the vibration of the object as a function of the time \cite{pedrini2006high,fu2007vibration}. The assessment  analysis of the motion can be done by phase difference or by Fourier analysis in the time domain. The method requires a CMOS camera, which can be costly.
%
It is also limited to low frequency vibrations, since a complete analysis of the motion requires a camera frame rate higher than the vibration frequency, because the bandwidth $\Delta \nu$ of the holographic signal, which is sampled  at the camera frame rate  $\nu_{CCD}$,  must be lower than  the Nyquist-Shannon  frequency limit $\Delta \nu<\nu_{CCD}/2$. For a periodic motion of  frequency $\nu_A$, the bandwidth $\Delta \nu \simeq 0$ is much smaller than $\nu_A$, and  $\nu_A$ can overcome this limit: $\nu_A>\nu_{CCD}/2$. The exposure time must remains nevertheless lower that the vibration period, in order to avoid to wash out the signal, whose phase may vary during  exposure. This  limits  the duty cycle and the thus SNR (Signal to Noise Ratio).

Powell and Stetson \cite{powell1965iva} have shown that the time averaged hologram of an harmonically vibrating object yields  yields alternate dark and bright fringes.
The dark fringes  correspond to the zeros of the Bessel function $J_0(z)$, where $z$ is proportional to the vibration amplitude. One gets then a direct mapping of this amplitude. Picard et al. \cite{picart2003tad} has simplified the processing of the data by performing  time averaged  holography   with a digital CCD camera. Time averaged  holography  has no limit in vibration frequency and do not involve costly laser system, nor an expensive CMOS fast camera. These advantages  yield numerous recent developments of the technique \cite{pinard2003musical,zhang2004vap,demoli2004detection,picart20052d,demoli2005dynamic,demoli2006real,asundi2006time,singh2007dynamic,picart2007tracking,fu2009vibration,kumar2009time}.
Although the time averaged method gives a way to determine the amplitude of vibration \cite{picart2005some}, quantitative measurement  remain quite difficult, especially for high vibration amplitudes. To solve this problem, Borza proposed high-resolution time-average speckle interferometry and digital holography techniques, which allow subpixel resolution (optical) phase recovery by inverting the Bessel function on its monotonic intervals and works well with larger vibration amplitudes \cite{borza2004high,borza2005mechanical,borza2006full}.

Joud et al. \cite{joud2008imaging}  extended the time averaged  technique to the detection of the optical signal at the  vibration sideband frequency $\nu_n$, where $n$ is the harmonic rank. It was demonstrated that the  signal amplitude of rank $n$ is then proportional to the Bessel function $J_n(z)$.
As in the seminal work of Aleksoff \cite{aleksoff1971temporally}, in Sideband holography  the reference beam  frequency is tuned in order to select the sideband of rank $n$. This tuning is made by the heterodyne holography technique \cite{le2000numerical}: the frequency of the illumination and  reference beams are shifted by acousto optic modulators (AOMs) so that the reference versus illumination frequency offset can be finely adjusted.
Sideband  holography, which is able to detect  MHz frequency vibrations \cite{gross2003shot},   is able on the other hand  to image sidebands up to rank $n=\pm 1000$ \cite{joud2009fringe}, and by the way to analyse vibration  of large   amplitude, which  can be studied by  time averaged  technique at the price of inverse methods \cite{borza2006full}.
%

Since both time averaged  and  sideband holography record the holographic signal over a large number a vibration periods, these two techniques are not sensitive to the phase of the vibration, and are thus unable to measure the instantaneous velocities of the object. To respond this problem, Leval et al. \cite{leval2005full} combine time averaged holography with stroboscopic illumination, but, since Leval uses a mechanical stroboscopic  device, the Leval technique suffer of a quite low duty cycle (1/144), and is limited in low vibration frequencies ($\nu_A<5$ kHz).

We propose here to combine Sideband Digital Holography (SDH) with stroboscopic illumination synchronized with the vibration.
To perform the stroboscopic illumination, we make use of  two AOMs  to control  both the frequencies and the amplitudes  of the  illumination and reference beams. This is done by  switching electronically on and off the Radio Frequency signals ($\simeq 80$ MHz) that drive the two AOMs.

If the amplitude of vibration is high, the light scattered by the vibrating object exhibit many sideband components. By sweeping the reference frequency such a way the holographic detection is tuned on the successive sideband harmonic ranks $n$,  stroboscopic sideband holography is then able to detect, for any stroboscopic time delay, all the harmonic ranks generated by Doppler effect. One can then  reconstruct the instantaneous velocity map of the vibrating  object. The mechanical phase, which is related to the sign of the velocity is obtained by the way. Note that since the stroboscopic illumination is made by  AOMs that are electronically driven, there is no practical limitation in the  stroboscopic frequency and  duty cycle.

In the following we will describe the principle of the method, and our optical and electronic setup. A test experiment is  made  with a clarinet reed excited sinusoidally  at $\nu_A=2$ kHz. The amplitude of vibration is such that we get sideband signal up to rank $n=\pm 100$. By sweeping the stroboscopic time delay, and by recording holograms for all harmonic ranks $n$, we can retrieve  images of the reed instantaneous velocity.

\section{Stroboscopic SDH principle}

\subsection{Periodic sinusoidal motion}

\begin{figure}[]
\begin{center}
  \includegraphics[width= 6.5 cm]{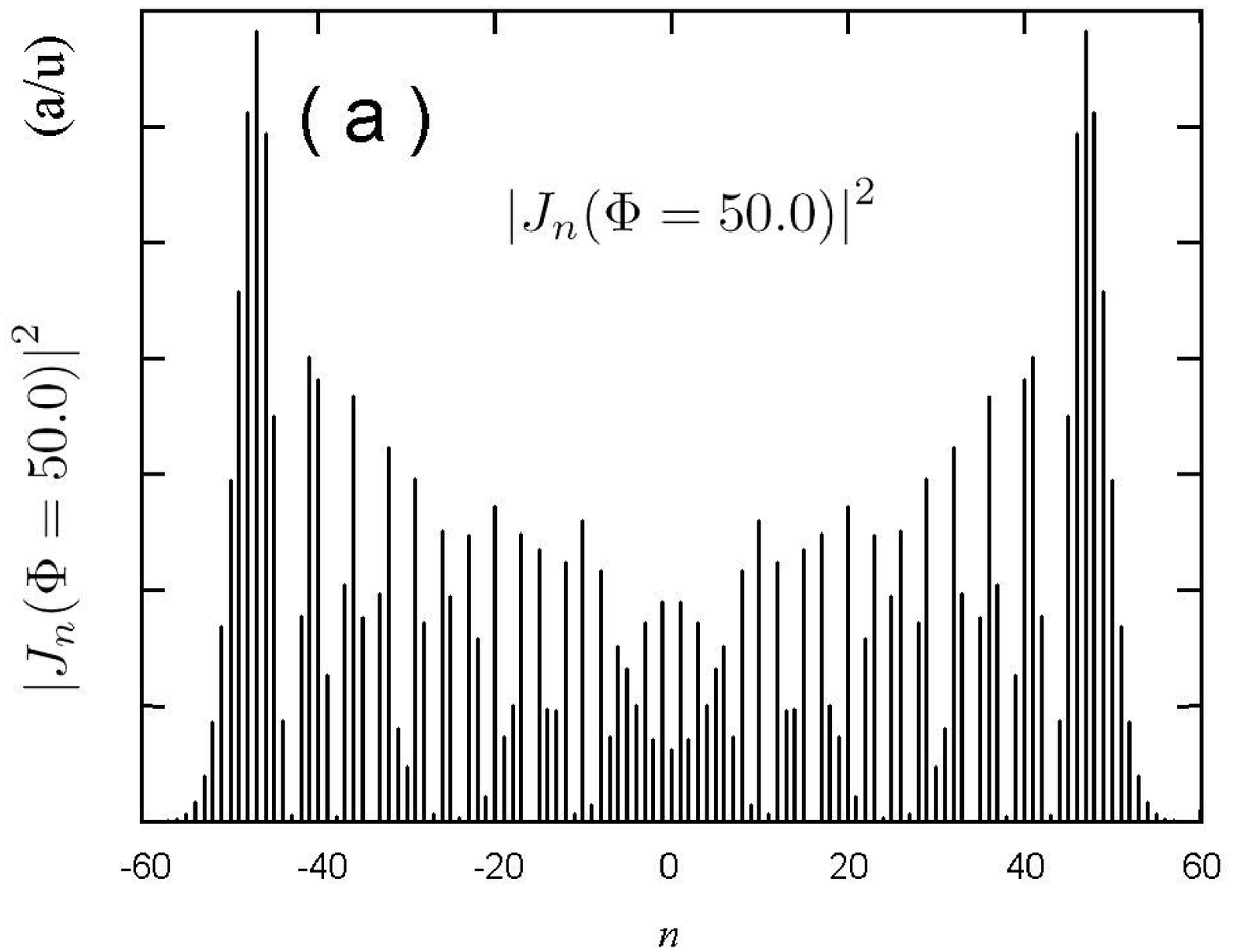}
 \includegraphics[width= 6.5 cm]{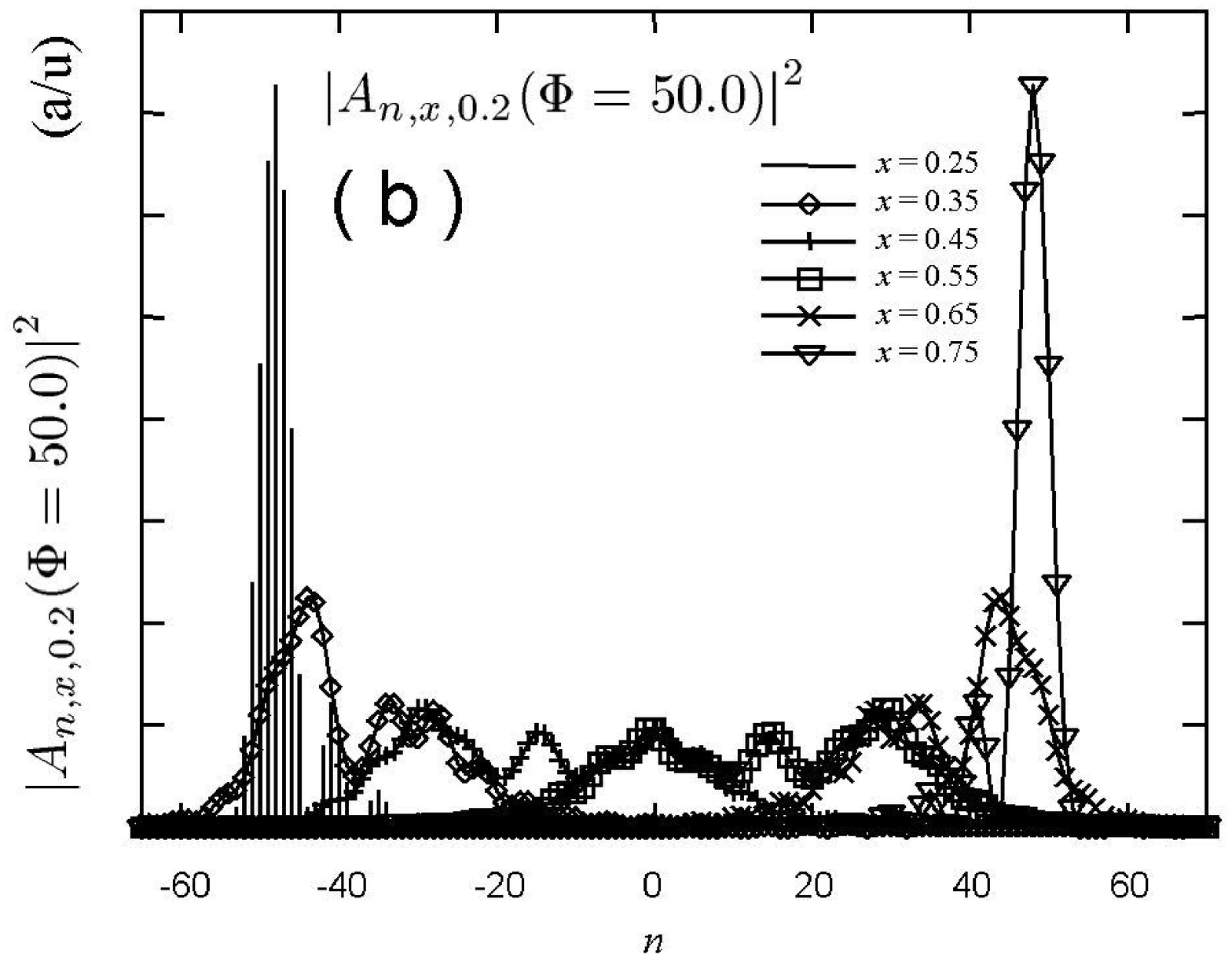}\\
  \includegraphics[width= 6.5 cm]{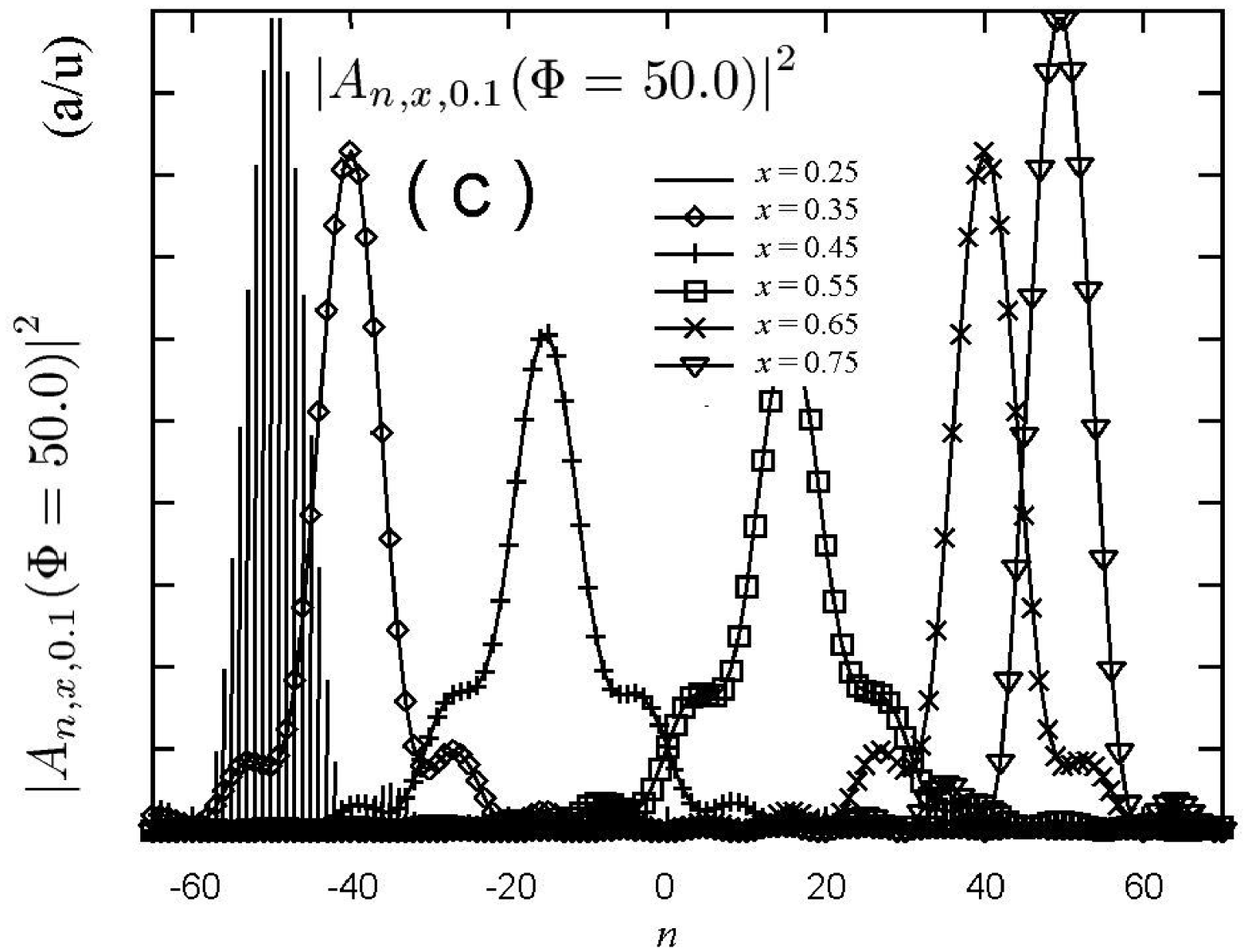}
 \includegraphics[width= 6.5 cm]{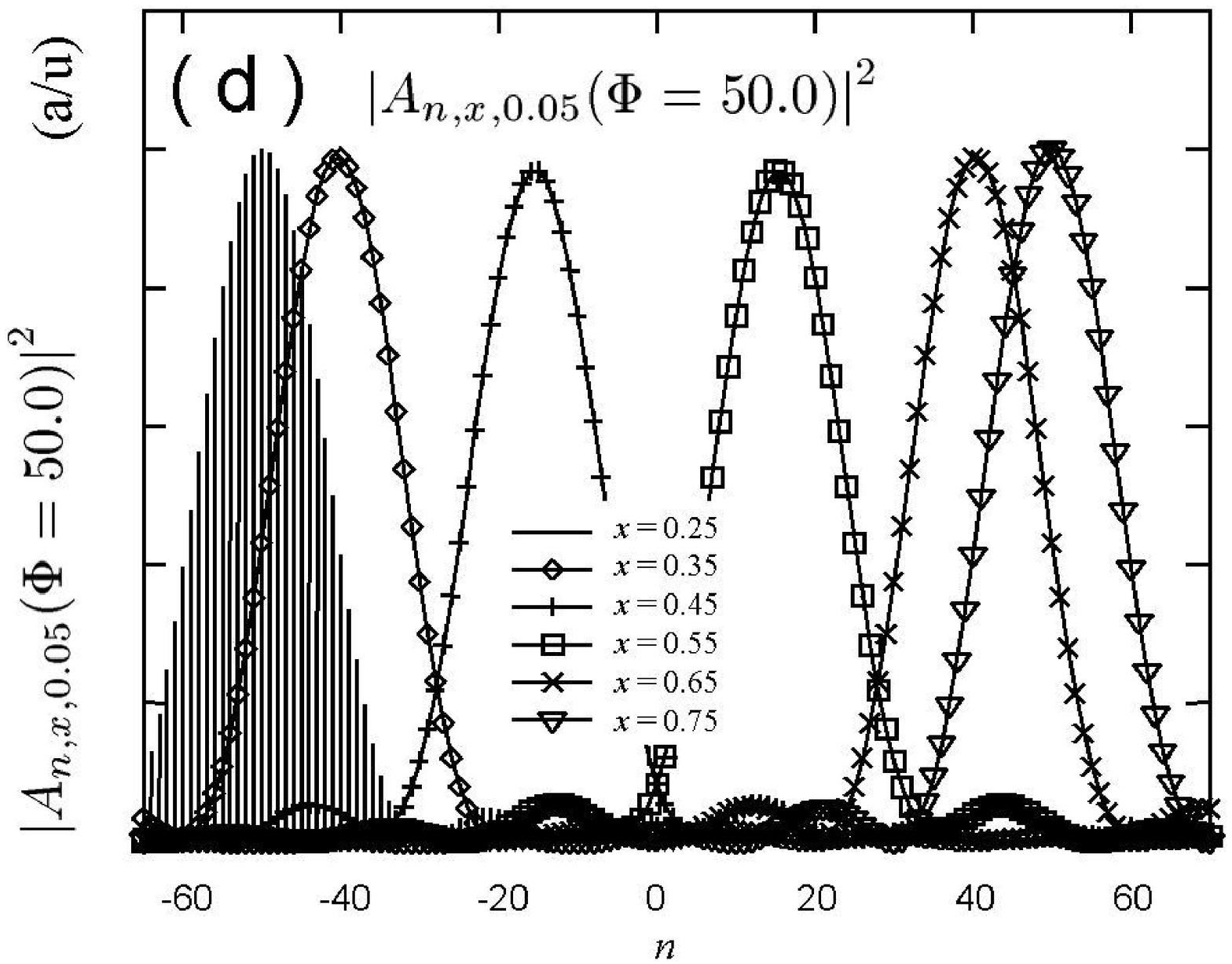}\\
  \caption{Distribution on the sidebands energy $|A_{n,x,\delta x}|^2$ as a function of the sideband harmonic $n$ without (a) and with (b-d) stroboscopic illumination. Curves are plotted for different stroboscopic illumination time $t=x T_A$ with $x=$ 0.25 0.35, 0.45, 0.55 , 0.65 and 0.75,  and different stroboscopic illumination duration $\Delta t=\delta x T_A$ with $\delta x=1$ (a), 0.2 (b), 0.1 (c) and 0.05 (d).  The vibration amplitude is $\Phi=50$.}\label{Fig_BesselJ}
\end{center}
\end{figure}

Consider a point of the object in vibrating sinusoidal at frequency
$\nu_A$ and amplitude $z_{max}$. Its displacement $z(t)$ is:

\begin{equation}\label{Eq_1}
  z(t) = z_{max} \cos(2\pi \nu_A t)
\end{equation}
In backscattering geometry, this corresponds to a phase modulation:
\begin{equation}\label{Eq_2}
   \varphi(t) = 4\pi z(t)/\lambda = \Phi \cos( 2\pi \nu_A t)
\end{equation}
where $\lambda$ is the optical wavelength and $\Phi =4\pi z_{max}/\lambda$. The scattered field is then
\begin{eqnarray}\label{Eq_3}
   E(t) &=& {\cal E} ~e^{j 2 \pi \nu_0 t+ j \varphi(t) }\\
   \nonumber &=& {\cal E} \sum_{n=-\infty}^{\infty} J_n(\Phi)e^{j 2 \pi( \nu_0 + n \nu_A)t }
\end{eqnarray}
where $\cal E$  is the complex amplitude of the field, $\nu_0$ the frequency of the illumination optical field  $E_I$,  and $J_n$ the n$^{\textrm{th}}$ order Bessel function of the first kind. The scattered field is a sum of components of frequency $\nu_n $:
\begin{eqnarray}\label{Eq_4}
  \nu_n=\nu_0 + n \nu_A
\end{eqnarray}
where $n$ is the  harmonic rank of the sideband ($n=0$ for the carrier). Equation \ref{Eq_3} means that the weight of the field component of frequency  $\nu_n$ is ${\cal E} J_n(\Phi)$.
Figure \ref{Fig_BesselJ} (a) shows how the energy of the sideband  $|{\cal E}|^2|J_n(\Phi)|^2$ varies with  $n$, assuming $\Phi=50$ rad.

To interpret the spectrum of  Fig. \ref{Fig_BesselJ} (a),  one can reverse Eq.\ref{Eq_4},  considering $n$ as a continuous variable related to the frequency $\nu \equiv \nu_n$ of the corresponding sideband component:
\begin{eqnarray}\label{Eq_5}
  n(\nu)=(\nu-\nu_0)/\nu_A
\end{eqnarray}
Here, $n$ is the Doppler frequency shift $\nu-\nu_0$ in $\nu_A$ units. This shift is, by definition, related to the distribution of the out-of-plane  velocity $V=dz/dt$ of the object:
\begin{eqnarray}\label{Eq_55}
 \nu-\nu_0= n \nu_A = 2  V / \lambda
\end{eqnarray}
where $\lambda$ is the optical wavelength.
The shift of harmonic rank  $n$  is confined between the values
$\pm \Phi $ that correspond to the maximum velocities $\pm V_{max}$ with $ V_{max} =   2\pi \nu_A  z_{max}$.
The discrete spectrum $|J_n(\phi=50)|^2$ of  Fig. \ref{Fig_BesselJ} (a) remains thus  mostly confined between  $\pm \Phi $, and drops abruptly from a maximum reached close to $n= \pm \Phi $ to almost zero \cite{joud2009fringe}.

In order to reconstruct the object velocity map at  given time of the vibration motion, we have considered a stroboscopic illumination. The field $E(t)$ is thus multiplied by the rectangular function $H_{x,\delta x}(t)$ of period $T_A=1/ \nu_A$
\begin{eqnarray}\label{Eq_6}
  H_{x,\delta x}(t) &=& 0  ~~\textrm{for}~~ t/T_A <x-{\delta x}/{2} \\
 \nonumber    &=& 1   ~~\textrm{for}~~ x-{\delta x}/{2} <t/T_A< x+{\delta x}/{2} \\
 \nonumber   &=& 0 ~~\textrm{for}~~ x+ {\delta x}/{2}< t/T_A
\end{eqnarray}
where $x T_A$ and $\delta x T_A$ are the illumination time and illumination duration.
The scattered field becomes thus:
\begin{eqnarray}\label{Eq_5_2}
   E(t) &=& {\cal E} ~H_{x,\delta x}(t)~e^{j 2\pi \nu_0 t+ j \varphi(t) }\\
\nonumber   &=& {\cal E} \sum_n A_{n,x,\delta x}(\Phi)   e^{j 2\pi \nu_n t}
\end{eqnarray}
where ${\cal E} A_{n,x,\delta x}$ is the amplitude of the n$^{\textrm{th}}$  sideband component.

We have calculated $A_{n,x,\delta x}$ from Eq.\ref{Eq_5_2} by numerical Fast Fourier Transform, and we have plotted on Fig. \ref{Fig_BesselJ} (b-d)   the energy  $|A_{n,x,\delta x}|^2$  as a function of  $n$  for different illumination phase delays  $x T_A$ and and different illumination durations $\delta x T_A$. The energy  $|A_{n,x,\delta x}|^2$  is centered on the Doppler shift $\overline{n}$ corresponding to the instantaneous velocity $V$ at  time $x T_A$:
\begin{eqnarray}\label{Eq_5_3}
   \overline{n} \nu_A = 2  \frac{v(x T_A)}{  \lambda}= -2  \frac{V_{max}}{\lambda} \sin(2\pi x )
\end{eqnarray}
For $\Phi=50$ and $x=$ 0.25 0.35, 0.45, 0.55 , 0.65 and 0.75, we get $\overline{n}=$ -50, -40.4, -15.4, +15.4  +40.4 and  +50 respectively, in good agrement with the curves plotted of Fig. \ref{Fig_BesselJ} (b-d). As expected, the shape of the energy spectrum  $|A_{n,x,\delta x}|^2$   strongly depends on the illumination duration $\delta x T_A$.

For short illumination duration (i.e. $\delta x=0.05$ on Fig. \ref{Fig_BesselJ} (d) ), we get a wide distribution of the energy along $n$, whose shape do  not depends on illumination time $x T_A$. This is expected since the illumination duration is to short to define the Doppler shift frequency precisely.  One is limited here  by a $\delta t \times \delta \nu \sim 1 $ ''Fourier  uncertainty principle'', where  $\delta t = \delta x T_A $ is the  width in time and  $\delta \nu = \delta n \nu_A $ the width in frequency ($\delta n$ being the width in harmonic rank).

For longer  illumination duration (i.e. $\delta x=0.2$ on Fig. \ref{Fig_BesselJ} (b) ),  the  shape of the energy spectrum $|A_{n,x,\delta x}|^2$ strongly depends on illumination time $x T_A$. It  can be narrow (for $x=0.25$ and 0.75) or wide (for $x=0.45$ and  0.55), since the  Doppler shift may vary slowly ($x=0.25$ and 0.75) or fast ($x=0.45$ and  0.55) during the illumination pulse. The best result (i.e. the narrower distribution of $n$ around  $\overline{n}$) is obtained for intermediate value of the illumination duration (i.e. for $\delta x=0.1$ on Fig. \ref{Fig_BesselJ} (c) ).

\subsection{Example of non sinusoidal motion: triangular motion}

\begin{figure}[]
\begin{center}
  \includegraphics[width= 6.5 cm]{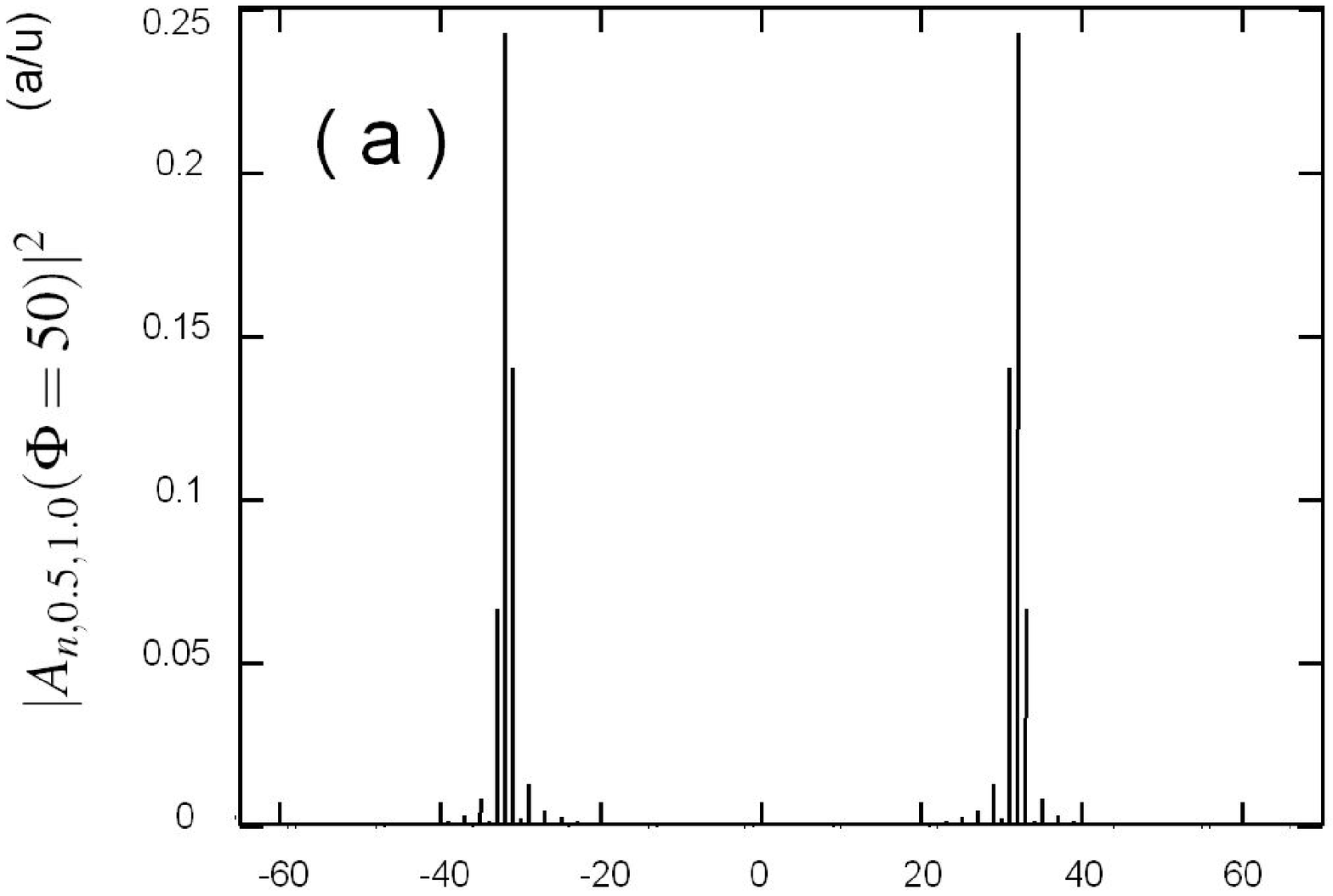}
 \includegraphics[width= 6.5 cm]{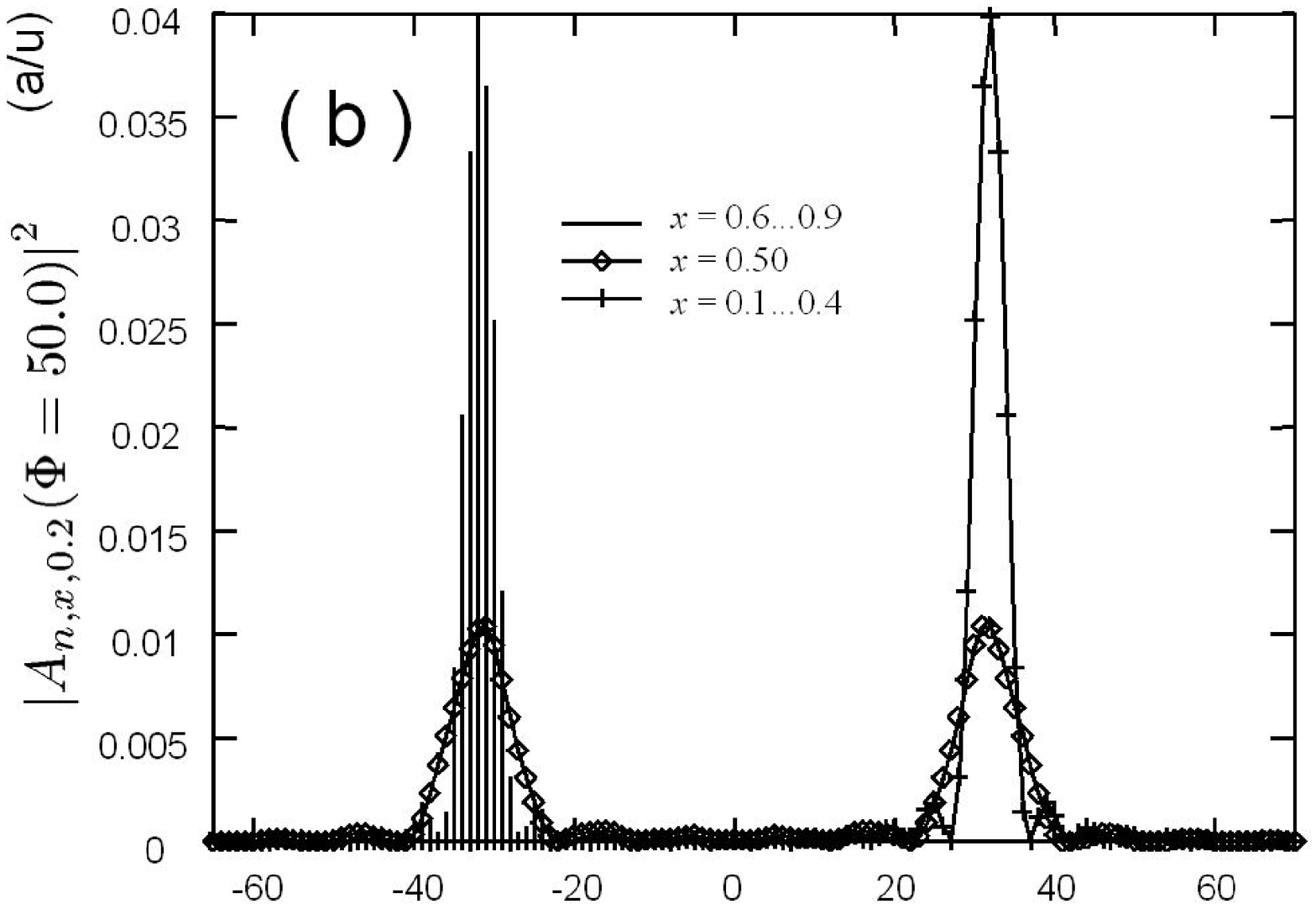}\\
  \includegraphics[width= 6.5 cm]{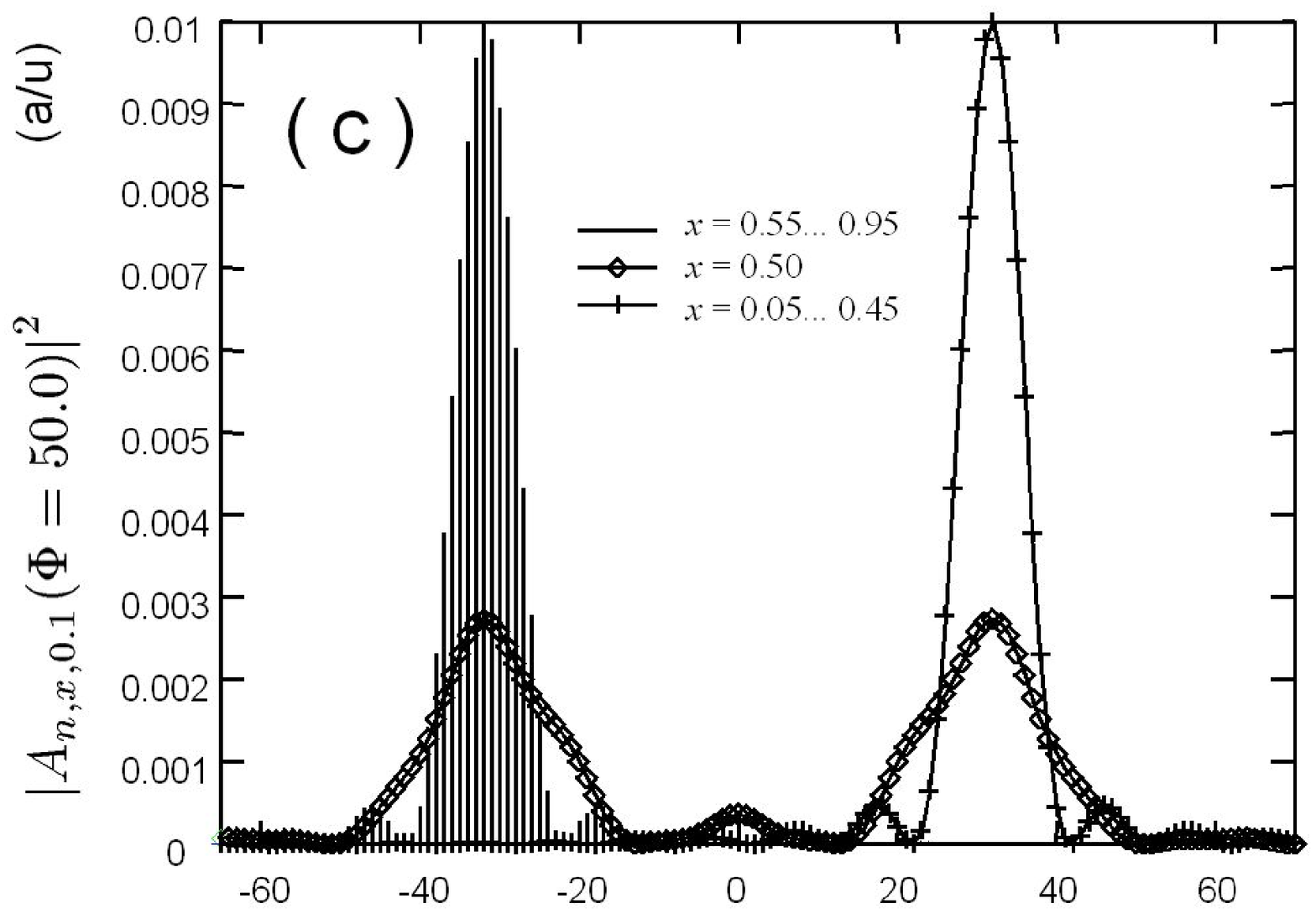}
 \includegraphics[width= 6.5 cm]{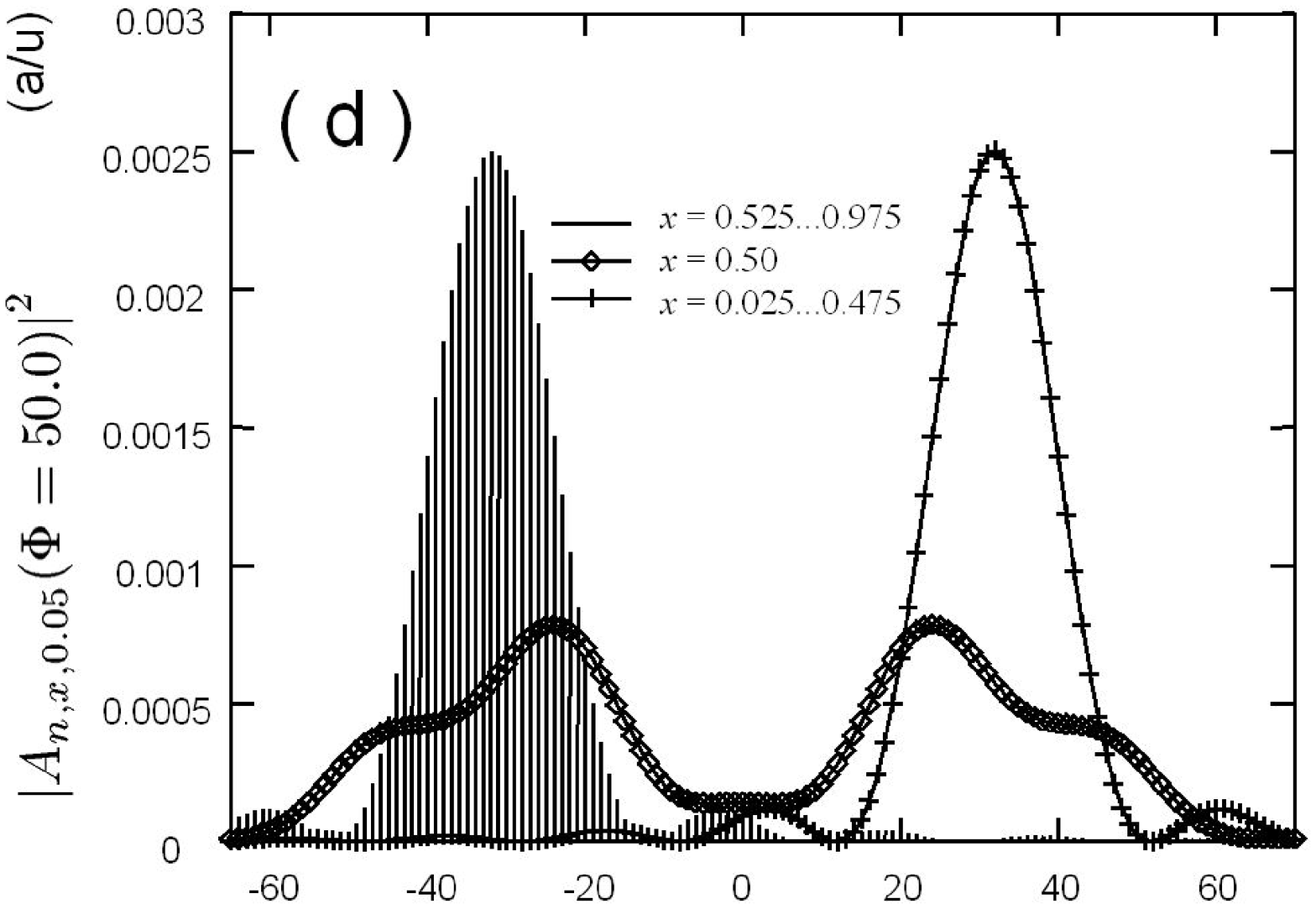}\\
  \caption{Distribution on the sidebands energy $|A_{n,x,\delta x}|^2$ as a function of the sideband harmonic $n$ without (a) and with (b-d) stroboscopic illumination for the triangular motion defined by Eq.\ref{Eq_100_2}. Curves are plotted for different stroboscopic illumination time $t=x T_A$,  and different stroboscopic illumination duration $\Delta t=\delta x T_A$ with $\delta x=1$ (a), 0.2 (b), 0.1 (c) and 0.05 (d).  The vibration maximum amplitude corresponds to $\Phi=50$.}\label{Fig_triangleJ}
\end{center}
\end{figure}

In the more general case of a non sinusoidal  but periodic motion of $z$ (period $T_A=1/\nu_A$), the calculation of the distribution of energy $|A_{n}|^2$ along the harmonic ranks $n$ can be made similarly. To illustrate the method, we have considered a triangular motion of period $T_A$: the displacement $z$  increases  from $z=-z_{max}$ at $t=0$ to $z=+z_{max}$ at $t=T_A/2$, and decreases from $z=+z_{max}$ at $t=T_A/2$ to $z=-z_{max}$ at $t=T_A$.

Equation \ref{Eq_5_2} remains  still valid for that motion, and can be used to calculate $A_{n,x,\delta x}$ by FFT as done for the sinusoidal motion. Nevertheless, the phase     $\varphi(t) = 4\pi z(t)/\lambda $ is no more sinusoidal, but  is given by:
\begin{eqnarray}\label{Eq_100_2}
  \varphi(t) &=&   (-1 + 4 t/ T_A )\Phi    \\
 \nonumber \varphi(t+T_A/2)  &=&  (+1 - 4  t/ T_A)\Phi
\end{eqnarray}
where  $0<t<T_A/2$ and  $\Phi =4\pi z_{max}/\lambda$.

We have calculated by FFT $A_{n,x,\delta x}$ from Eq.\ref{Eq_5_2} and  Eq.\ref{Eq_100_2}.  We have plotted the energy  distribution $|A_{n,x,\delta x}|^2$ without stroboscopic illumination  on Fig. \ref{Fig_triangleJ} (a), and  with stroboscopic illumination on   Fig. \ref{Fig_triangleJ} (b-d). We have plotted  $|A_{n,x,\delta x}|^2$    for different illumination phase delays  $x T_A$ and and different illumination durations $\delta x T_A$. The distributions  of the energy  $|A_{n,x,\delta x}|^2$ plotted here on Fig. \ref{Fig_triangleJ} are very different than for the sinusoidal motion of same maximum amplitude $\pm z_{max}$ plotted  on Fig. \ref{Fig_BesselJ}.

For  $ 0.5 <x<1.0$, the displacement $z$ decreases at constant velocity, and the shape of the energy distribution $|A_{n,x,\delta x}|^2$, which  is shifted  to the  negative harmonic rank $n<0$,  not depends on the illumination time $x$.  For  $ 0 <x<0.5$ a similar result is obtained:  the shape do not change, but the shift is positive: $n>0$ since $z$ increases. These results are  expected for a triangular motion.

The analysis of distribution of energy $|A_{n,x, \delta x}|^2$ as a function of harmonic rank $n$, illumination time $x T_A$ and illumination duration $\delta x T_A$ yields detailed informations on the motion and on its non harmonic components.

\section{Experimental  setup}

\begin{figure}
\begin{center}
  \includegraphics[width=8 cm]{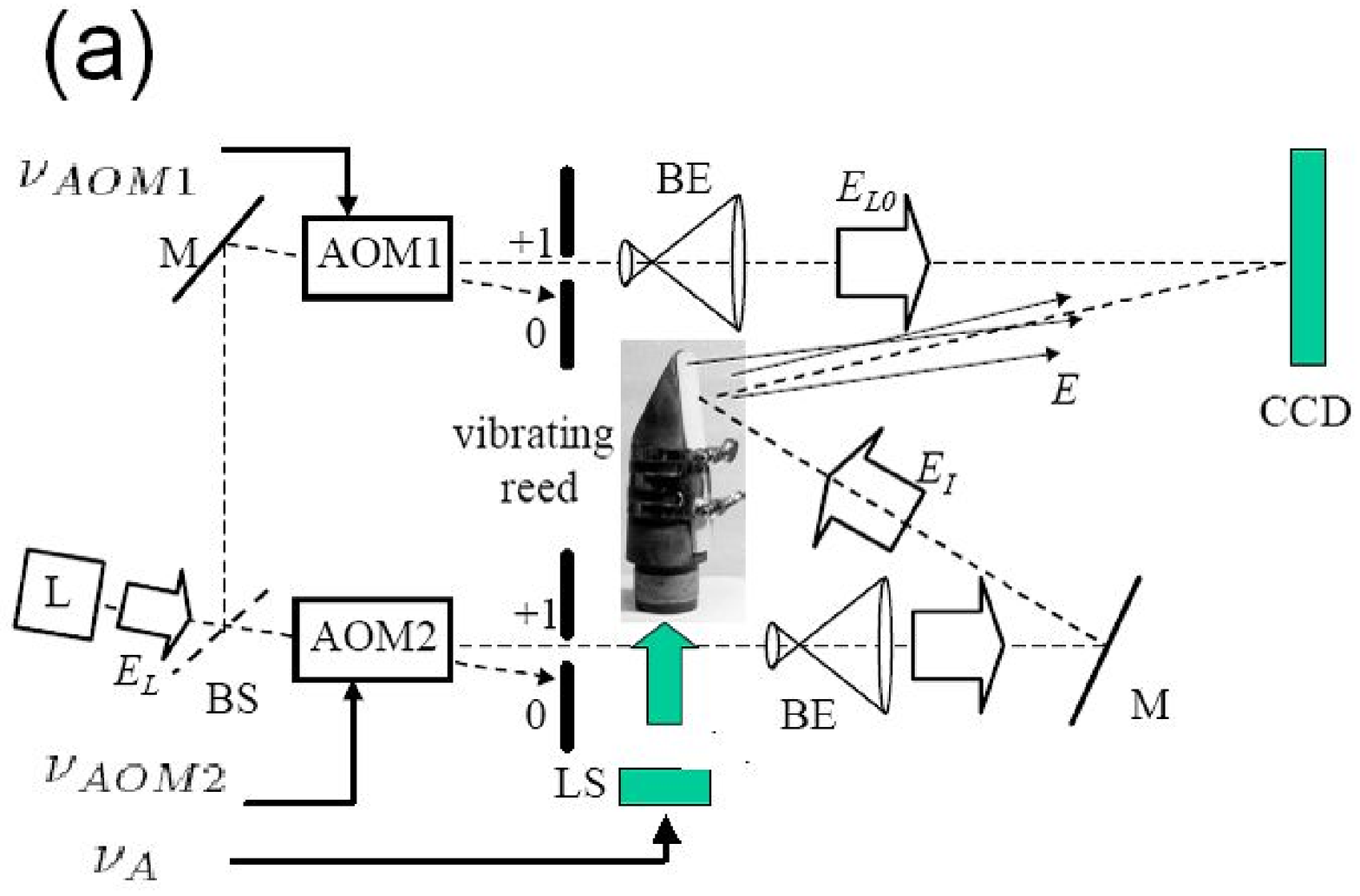}\\
  \includegraphics[width=8 cm]{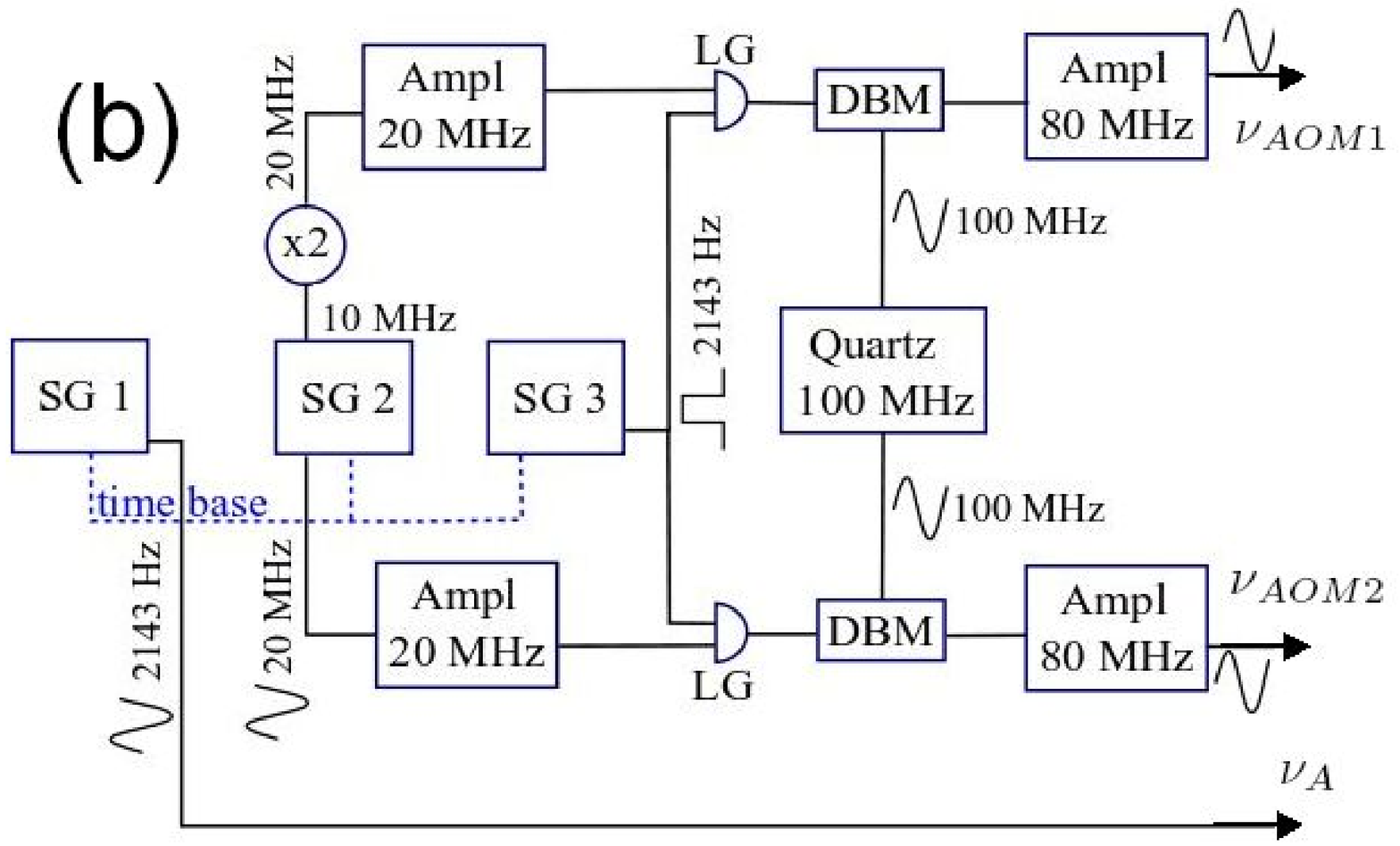}\\
  \caption{Optical setup (a), and  electronic (b) that drives the acousto optic modulator and the loudspeeker. (a)    L: laser;  AOM1,2: accousto optic modulators driven at frequency $\nu_{AOM1,2}$;  M: mirrors;  BS:  beam splitter; BE:  beam expander; LS: loudspeaker exciting the clarinet reed at frequency $\nu_A/2$;  CCD:  CCD camera. (b)  SG1,2,3:  direct digital synthesizer (DDS) signal generators;; Quartz: 100 MHz quartz oscillator;  LG: NAND logical gate; DBM:  double balanced mixer; Ampl: 20 MHz and 80 MHz LC resonant  amplifiers.}\label{Fig_setup}
\end{center}
\end{figure}

The sideband holography setups presented in  previous publications  does not allow to reconstruct the object at a given time of the vibration motion. To circumvent this problem, we have   realized a stroboscopic illumination synchronized on the vibration motion.
Figure \ref{Fig_setup}  (a) shows
%
%
the optical part of our setup, which is   similar to the  setups  previously described \cite{joud2008imaging,joud2009fringe}.
Figure \ref{Fig_setup}  (b) shows  the electronic, which  drives the  acousto optics modulators AOM1 and AOM2, and which  has been modified  to add amplitude modulation abilities.
This electronic is based on 3 direct digital synthesizer  signal generators  (SG1, 2 and 3), which are locked on  the same 10 MHz time base.
SG1 excites the loudspeaker, which makes  the reed vibrate at $\nu_A=$ 2143 Hz.
SG2 generates  a frequency tunable sinusoidal signal at $\nu_{SG2} \simeq 20$ MHz, and a reference fixed  frequency 10 MHz signal, which is doubled by a  frequency doubler (FD2: Mini Circuit Lab Inc.).
SG3 generates a  rectangular gate of duration $0.15 \times T_A = 70~ \mu \textrm{s}$ at frequency $\nu_A$. This gate (SG3) is synchronized with respect to the reed excitation (SG1) with an electronically adjustable phase delay $x T_A$.

To realize the stroboscopic illumination, two NAND Logical Gates LG, driven by the SG3 gate,  switch on and off  the adjustable and the fixed 20 MHz frequency signals.
The two 20 MHz  gated  signals obtained by the way are mixed with the signal of a 100 MHz quartz oscillator by two doubled balanced mixer (ZAD-1H: Mini Circuit Lab Inc.). We get thus two gated signals that exhibit two frequency components  $\simeq$ 100 MHz $\pm$ 20 MHz. The signals at
$\nu_{AOM1}= 80 $ MHz and $\nu_{AOM2}\simeq  80 $ MHz that drive the acousto optic modulators  AOM1 and AOM2, are thus obtained by filtering the  $\simeq$ 100 MHz $\pm$ 20 MHz signals with two 80 MHz LC resonant  amplifiers.

\begin{figure}
\begin{center}
  \includegraphics[width=10 cm]{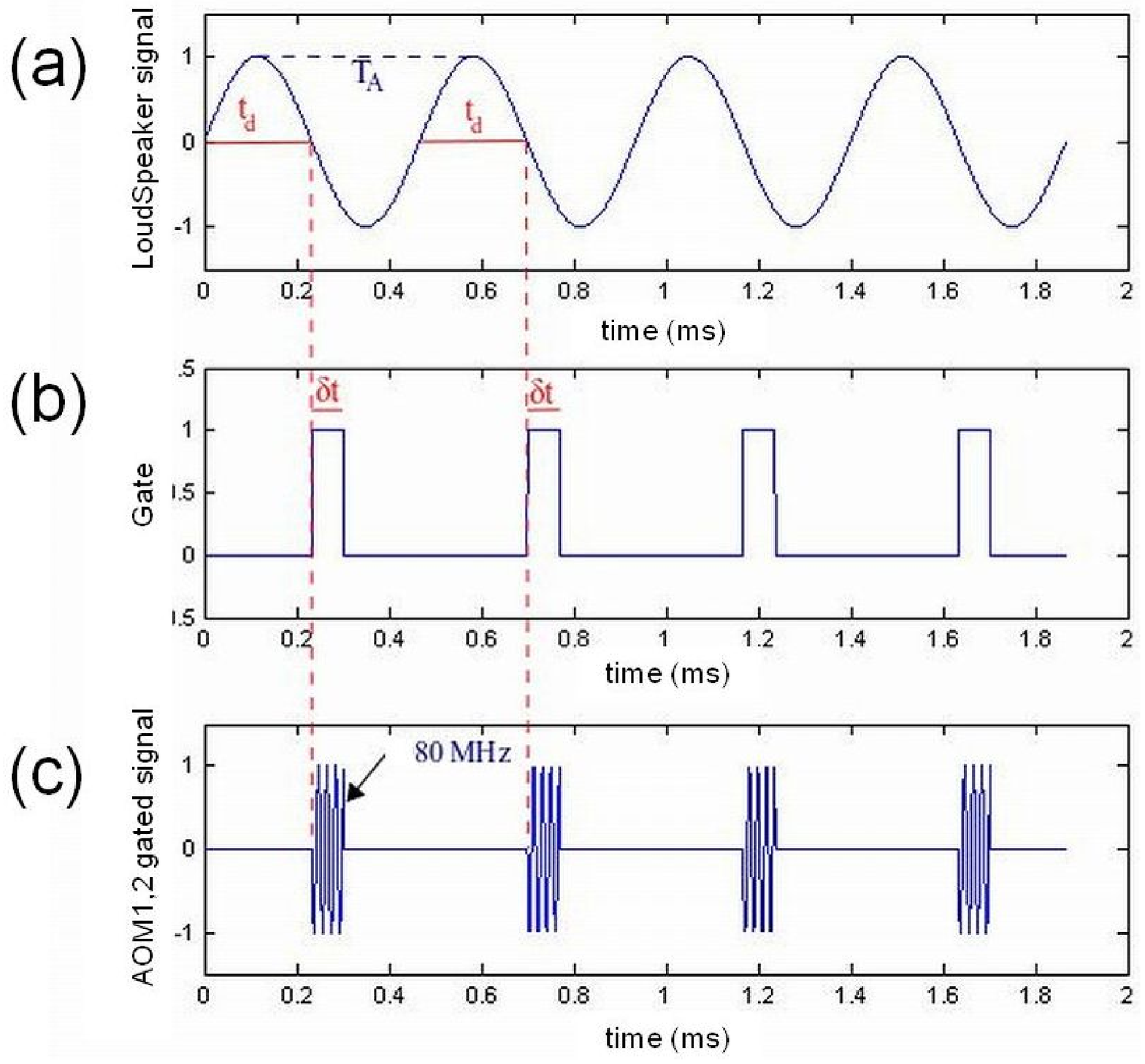}
  \caption{  Chronogram of the signals: (a) SG1 sinusoidal signal of period  $T_A=1/\nu_A$ exciting the reed; (b) SG3 rectangular  gate and  (c) $\simeq$80 MHz gated signals driving the accousto-optical modulators AOM1 and AOM2. }\label{Fig_chronogram}
\end{center}
\end{figure}

Figure \ref{Fig_chronogram} illustrates the synchronization of the various signals. Out of the gate, the $\nu_{AOM2}$ and $\nu_{AOM2}$  signals are off, and one get no optical signals on the  grating orders +1 of the acousto optic modulators AOM1 and AOM2. The illumination and reference beams are then off. Our stroboscopic gate acts thus both on illumination, and on holographic reference (i.e. on holographic detection efficiency).

The data acquisition is made by a computer that drive the three signal generators SG1, SG2 and SG3. For every value of the stroboscopic delay  $x T_A$,  the harmonic rank of detection is  swept from  $n =-100$ to  $n = +100$, i.e., for each $n$,  the computer drives SG2 such a way  the frequencies of the acousto-optics modulators  signals verify:
\begin{eqnarray}\label{Eq_7}
 \nu_{AOM1} -\nu_{AOM2}= n \nu_{A} +  \nu_{CCD}/4
\end{eqnarray}
Four-phase heterodyne holography is then performed \cite{le2000numerical}, i.e., a sequence of 4 images $I_0, ..., I_3$ is  recorded yielding  the complex hologram:
\begin{eqnarray}\label{Eq_8}
H= (I_0-I_2)+ j (I_0-I_2)
\end{eqnarray}
The optical field image of the reed $E(x,y)$ is  reconstructed from  $H(x,y)$ with the standard one FFT (Fast Fourier Transform) method \cite{schnars1994direct,picart2008general}, where $x$ and $y$ are the coordinates of the $1024 \times 1024$ calculation matrix. The reconstructed images intensity $I=|E|^2$ of rank $n$ is  stored in a 3D $1024\times 1024\times 201$ cube of data with axes  $x$, $y$ and $n$.  To the end, an image of the reed velocities is  extracted  from the cube.

For each stroboscopic delay $xT_A$ the operations are repeated:  the rank $n$ is swept; for each $n$  the holograms are recorded, reconstructed  and stored within  the cube of data. Images of the  velocities  are then  obtained.

\section{Experimental results}

\begin{figure}[]
\begin{center}
  \includegraphics[width=4.2 cm]{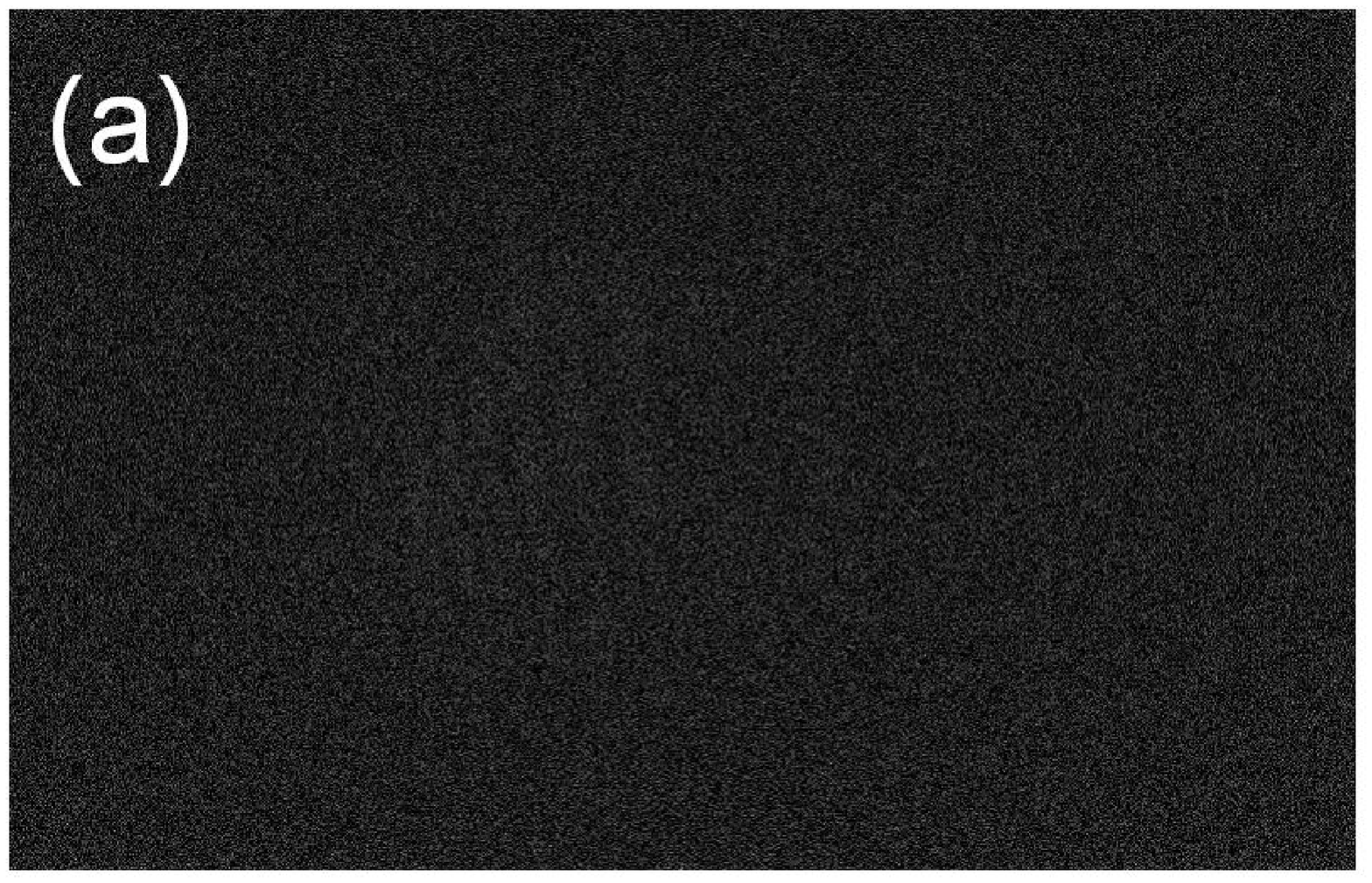}
  \includegraphics[width=4.2 cm]{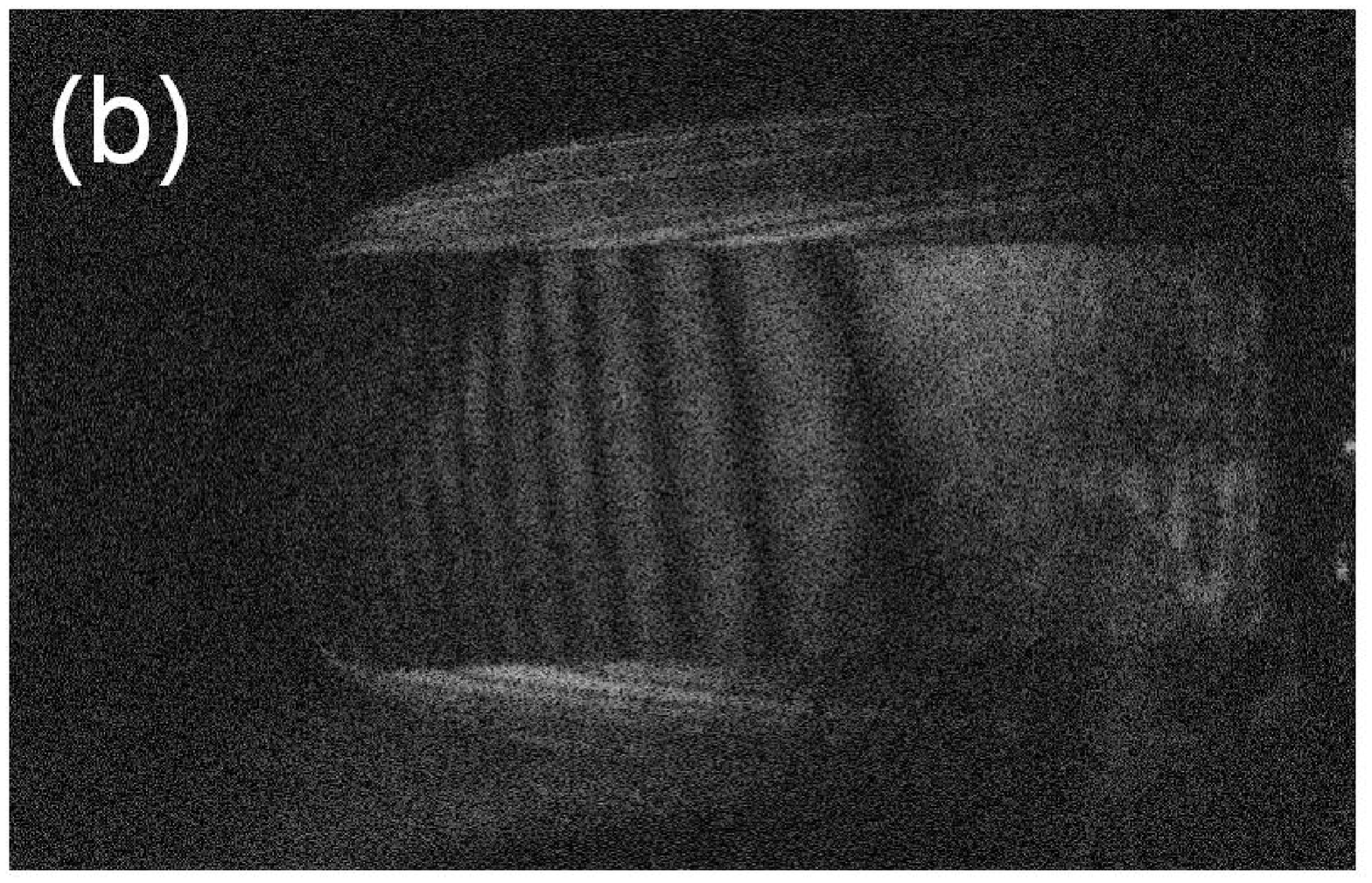}
  \includegraphics[width=4.2 cm]{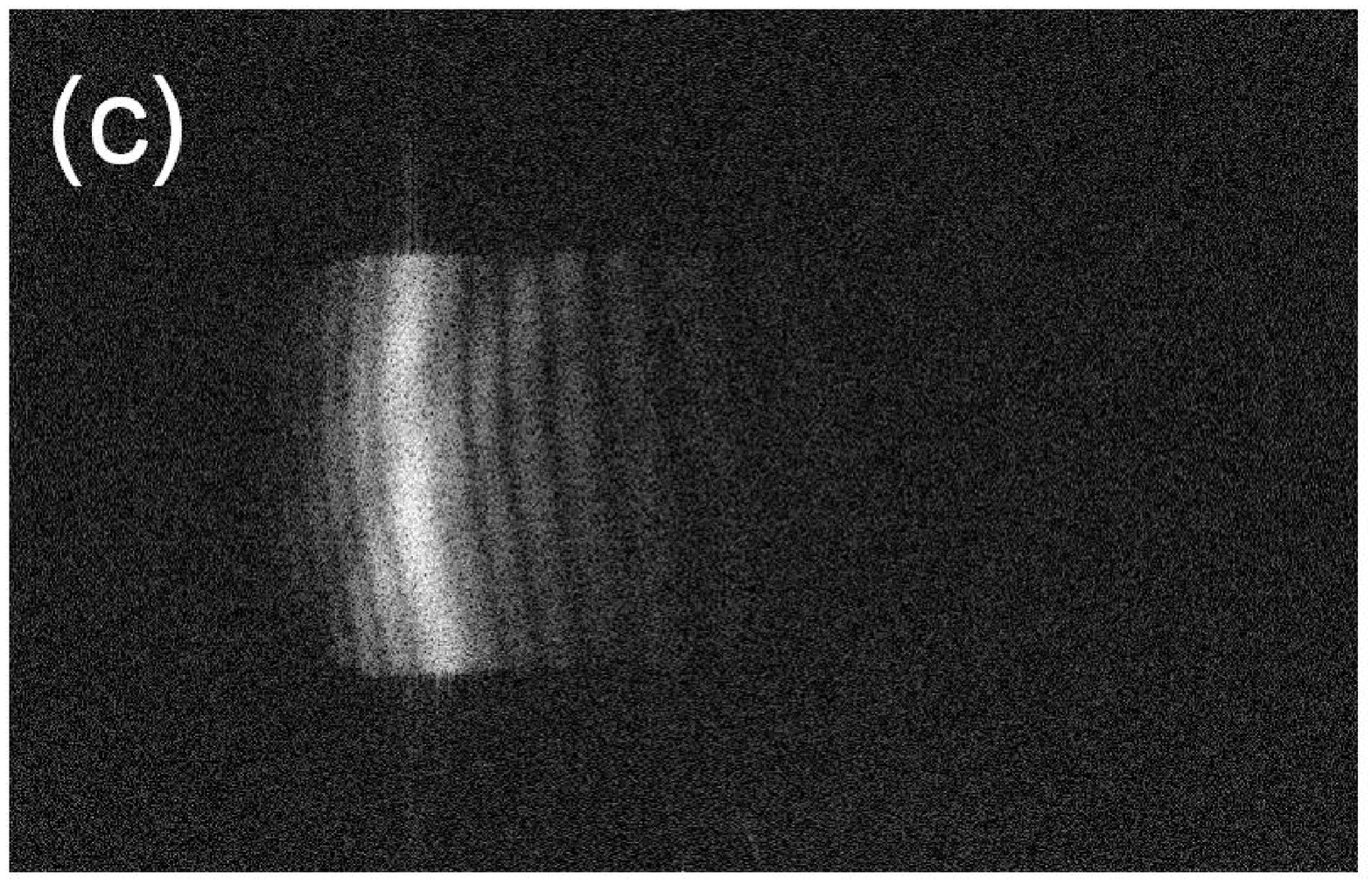}\\
  \includegraphics[width=4.2 cm]{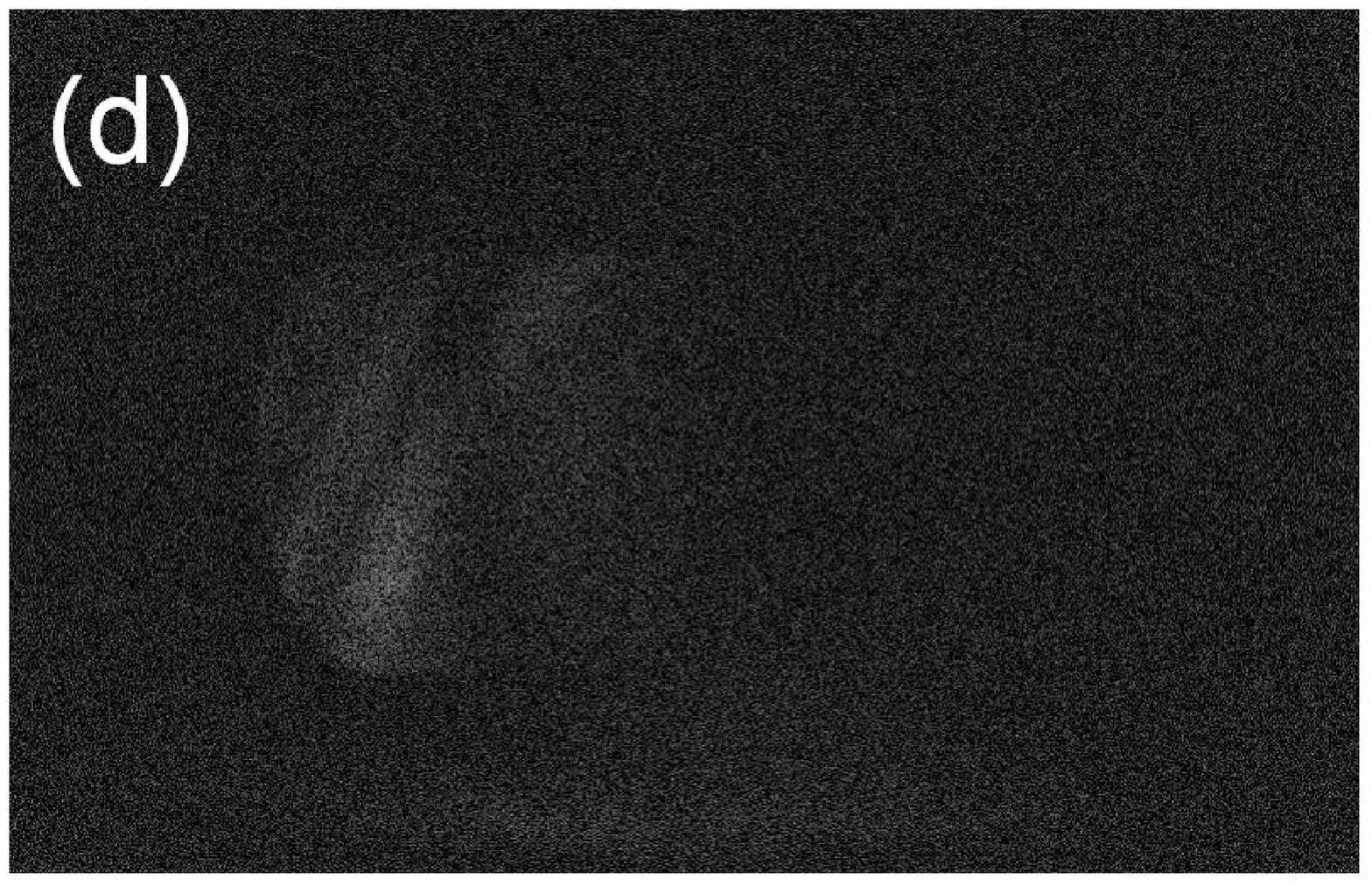}
  \includegraphics[width=4.2 cm]{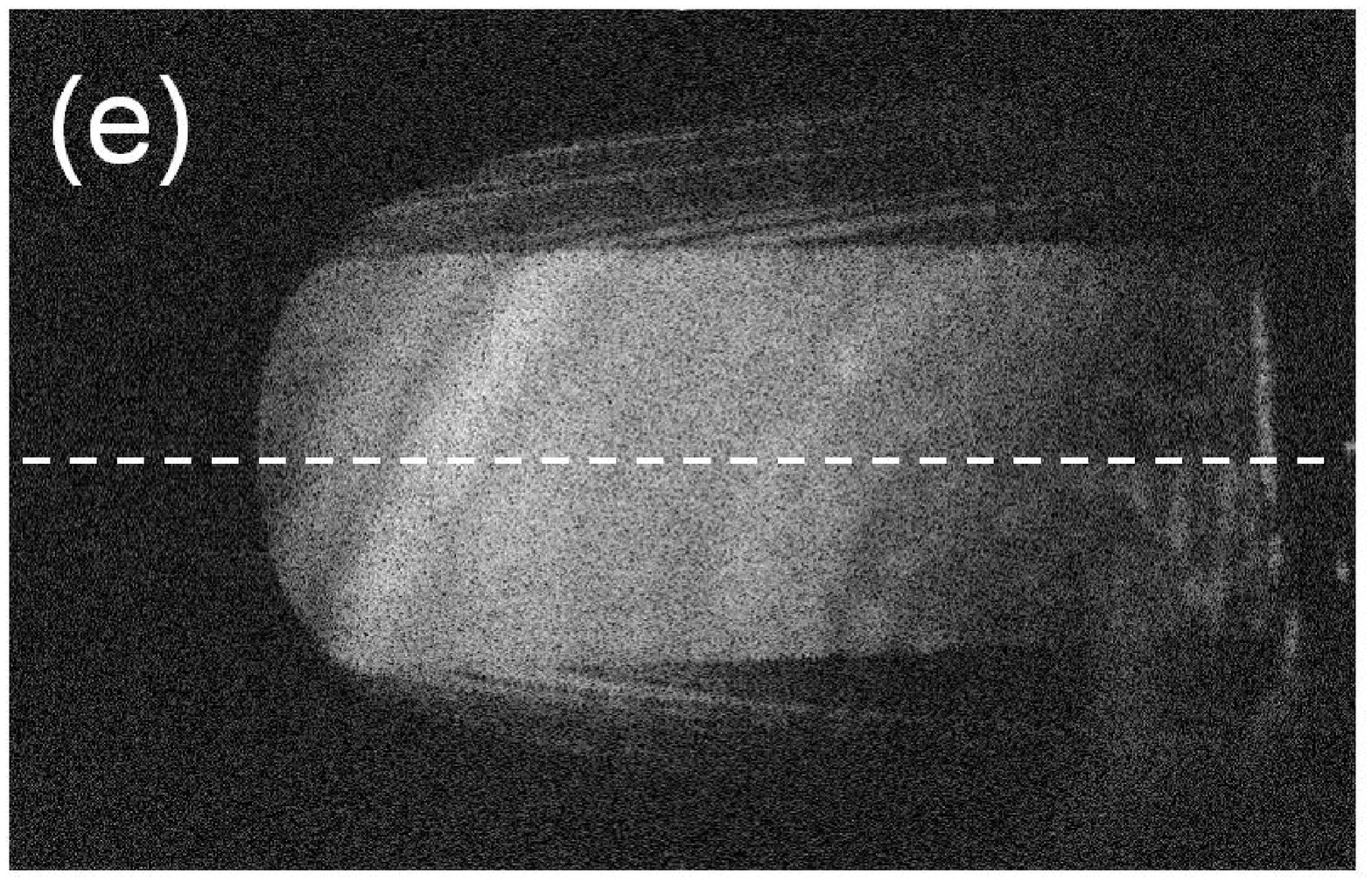}
  \includegraphics[width=4.2 cm]{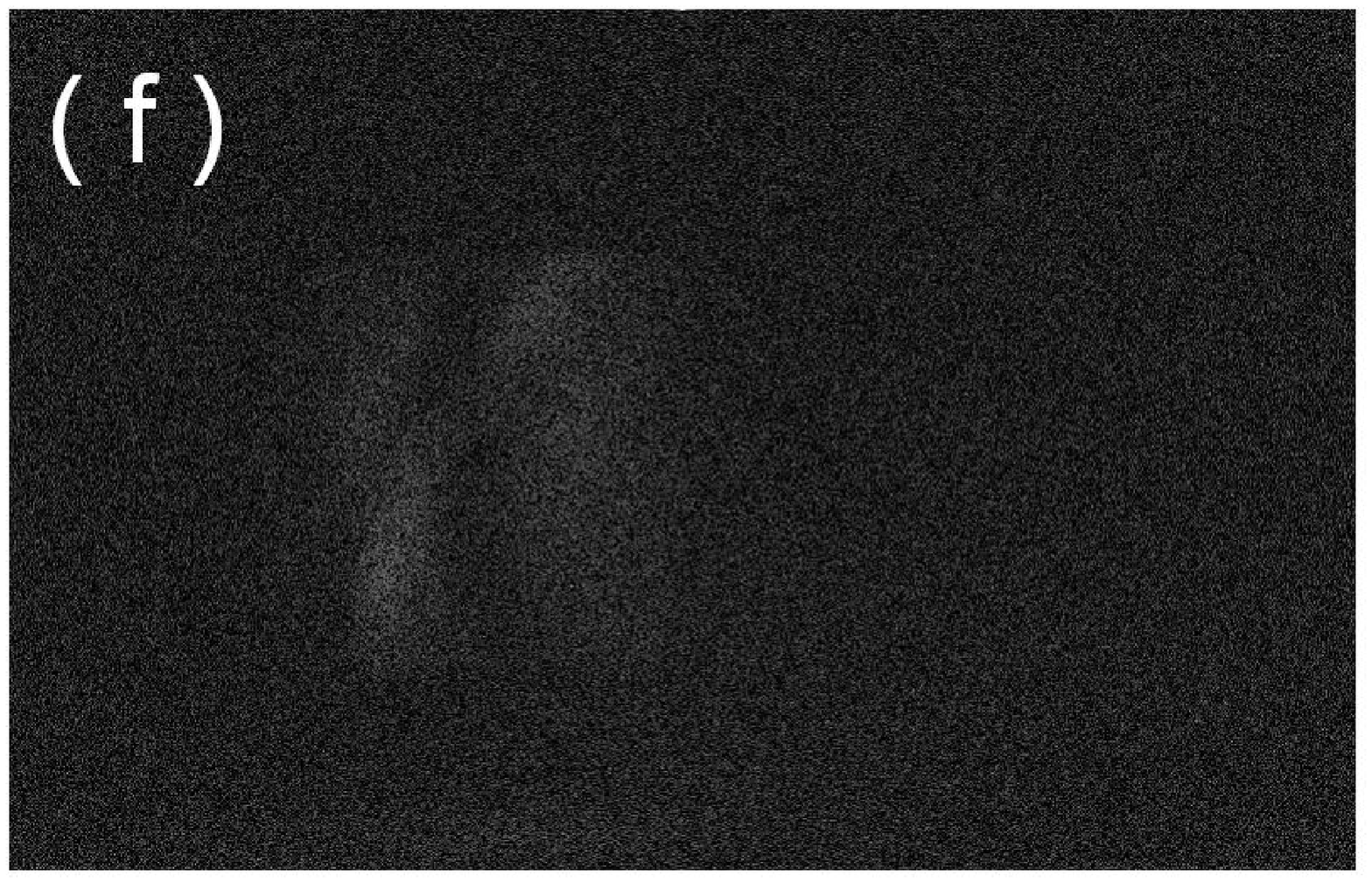}\\
  \includegraphics[width=4.2 cm]{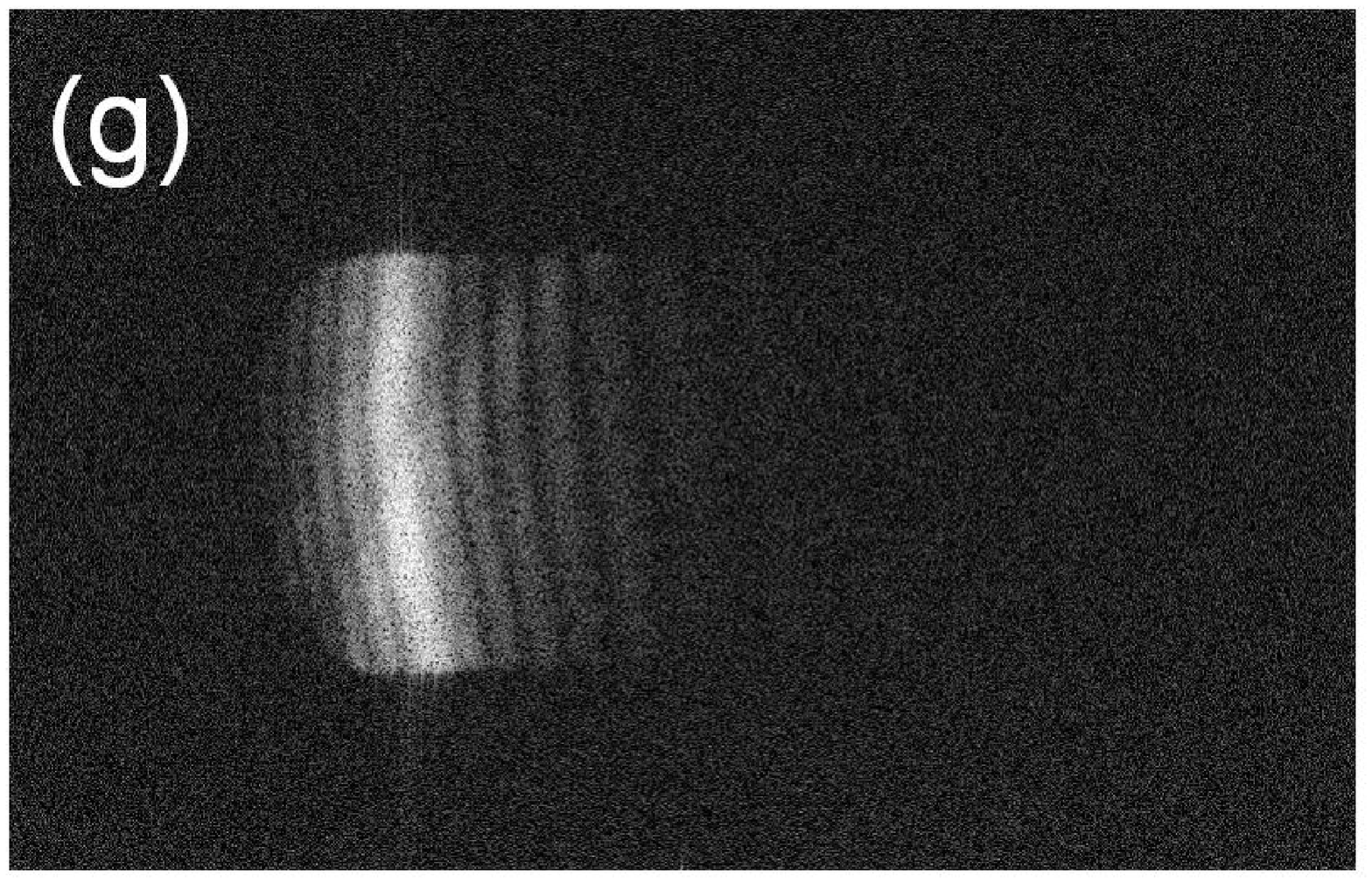}
    \includegraphics[width=4.2 cm]{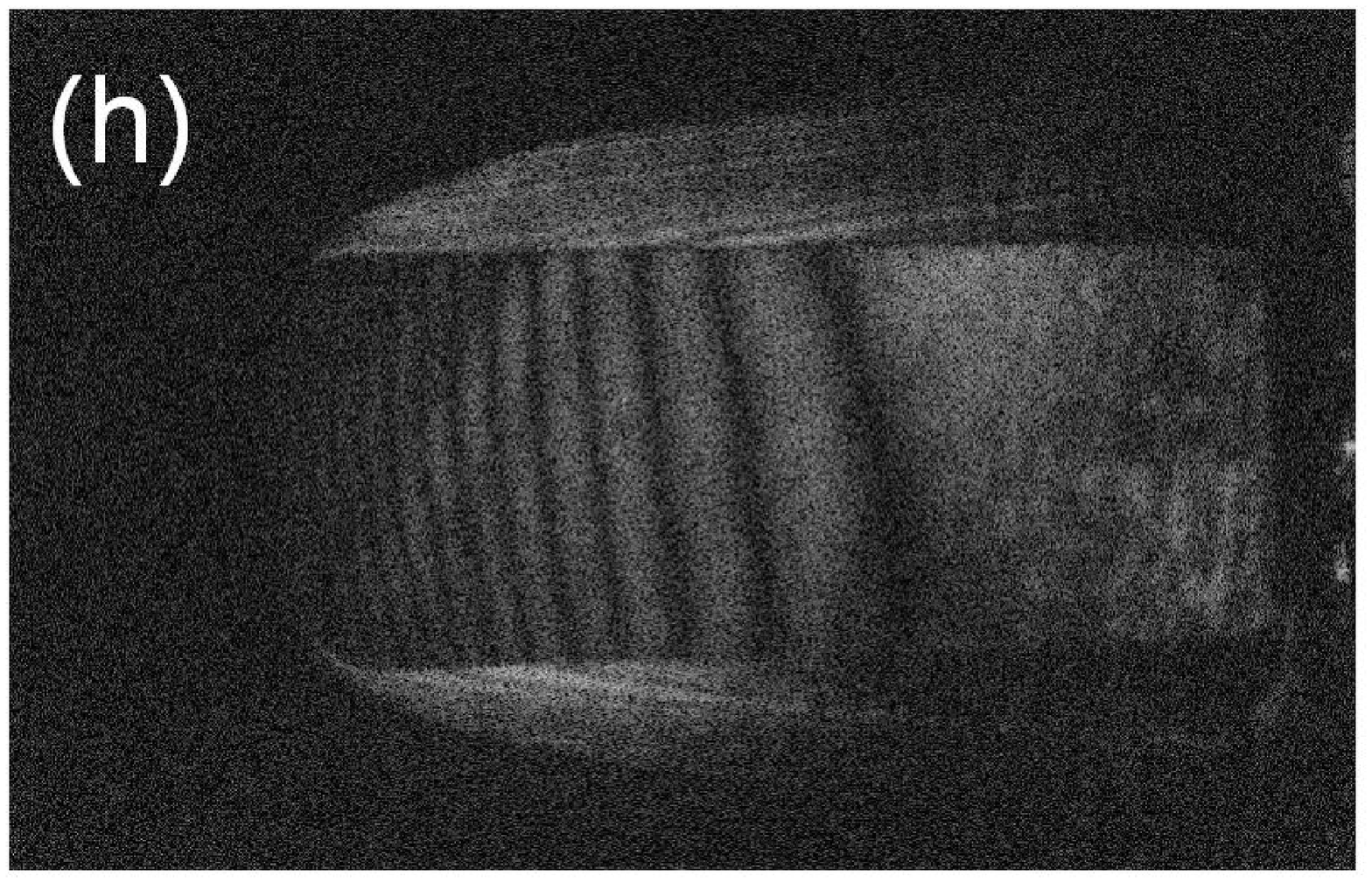}
  \includegraphics[width=4.2 cm]{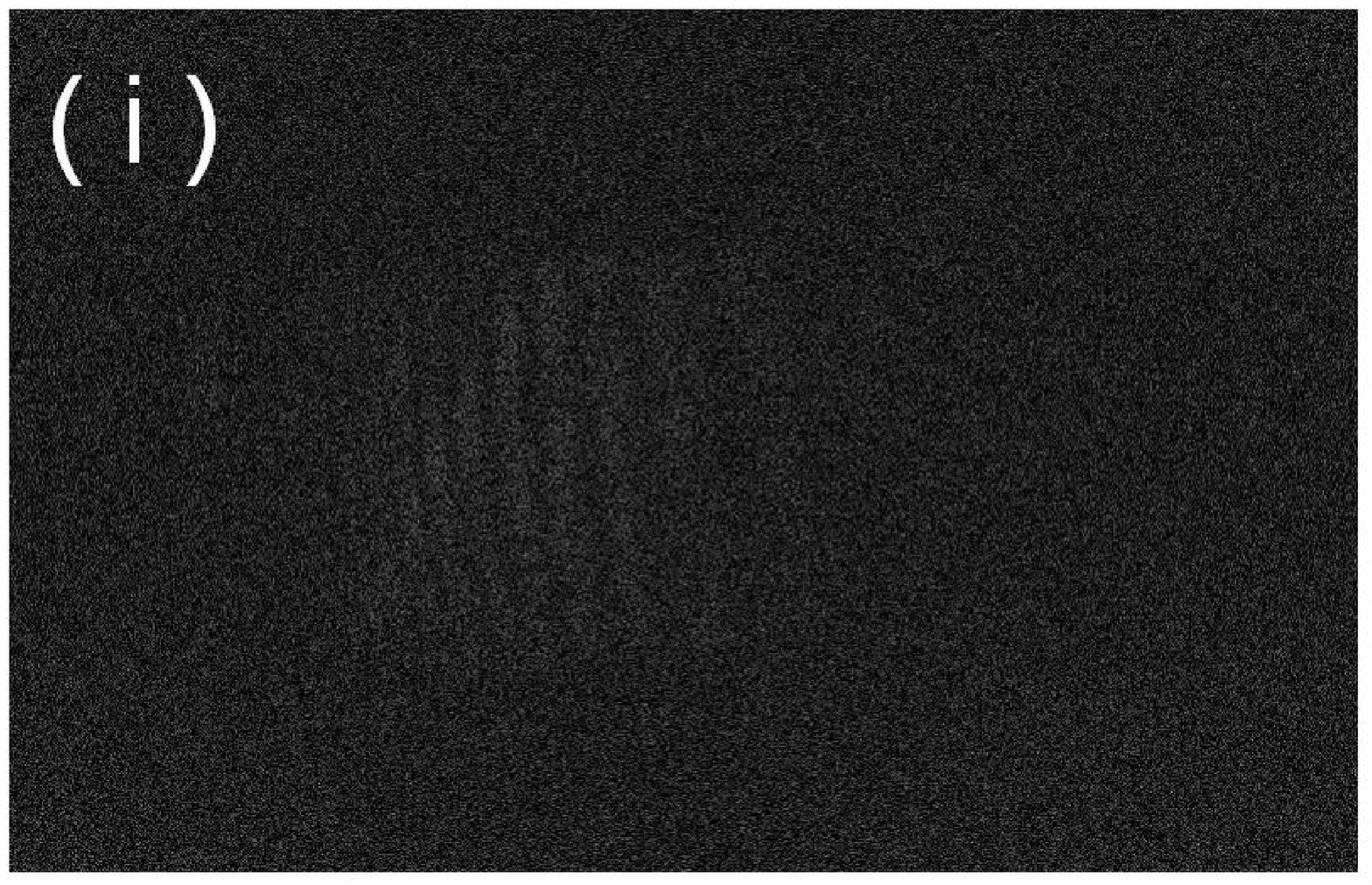}\\
  \caption{Reconstruction images of the reed  at time $x T_A = 0$ (a,b,c), $0.3~ T_A$ (d,e,f)    and $0.7 ~T_A$ (g,h,i)  for $n =  -60$ (a,d,g),  $0$ (b,e,h) and $+60.0$ (c,f,i). The images are displayed in logarithmic scale for the field intensity $|E|^2$.}\label{Fig_images_log_xy_strobo}
\end{center}
\end{figure}

\begin{figure}[]
\begin{center}
\includegraphics[width=7 cm]{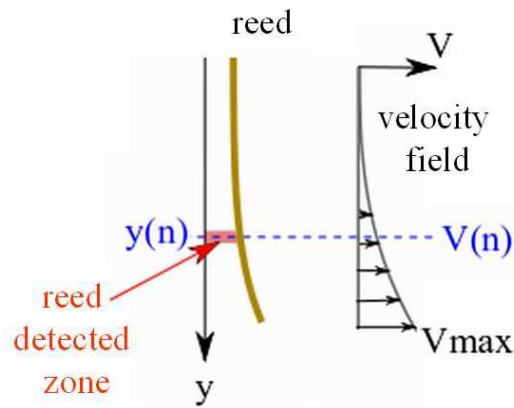}\\
  \caption{Reed displacement $z(y,xT_A)$ and the reed velocity $V(y,xT_A)$ at illumination time $t=x T_A$. The holographic signal on sideband of rank $n$ is obtained  if $V(y,xT_A)$  is close to $V(n)= n \lambda \nu_A$.  }\label{Fig_velocity_field}
\end{center}
\end{figure}

Figure \ref{Fig_images_log_xy_strobo} shows nine reconstructed   images  of the reed with stroboscopic delay $x T_A$ with $x=0$, $ 0.3$  and $0.7 $ (respectively from top to bottom), and for $n = -60$, $0$ and $+60$ (respectively from left to right). Each image of  Fig. \ref{Fig_images_log_xy_strobo}  represents thus one   component $n$ of the instantaneous Doppler  spectrum.
Since the hologram are recorded off axis, the read images are off axis too. To center the images of the reed, the $1024\times 1024$ reconstructed images are truncated to $1024\times 512$ on Fig.  \ref{Fig_images_log_xy_strobo}. One recognize the reed on Fig.  \ref{Fig_images_log_xy_strobo} (b) to (h), and the reed + mouthpiece on Fig.  \ref{Fig_images_log_xy_strobo} (b,e and h). The right part of reed is clamped on the mouthpiece, while the tip, whose motion is free, is on the left.

\begin{figure}[]
\begin{center}
  \includegraphics[width=13 cm]{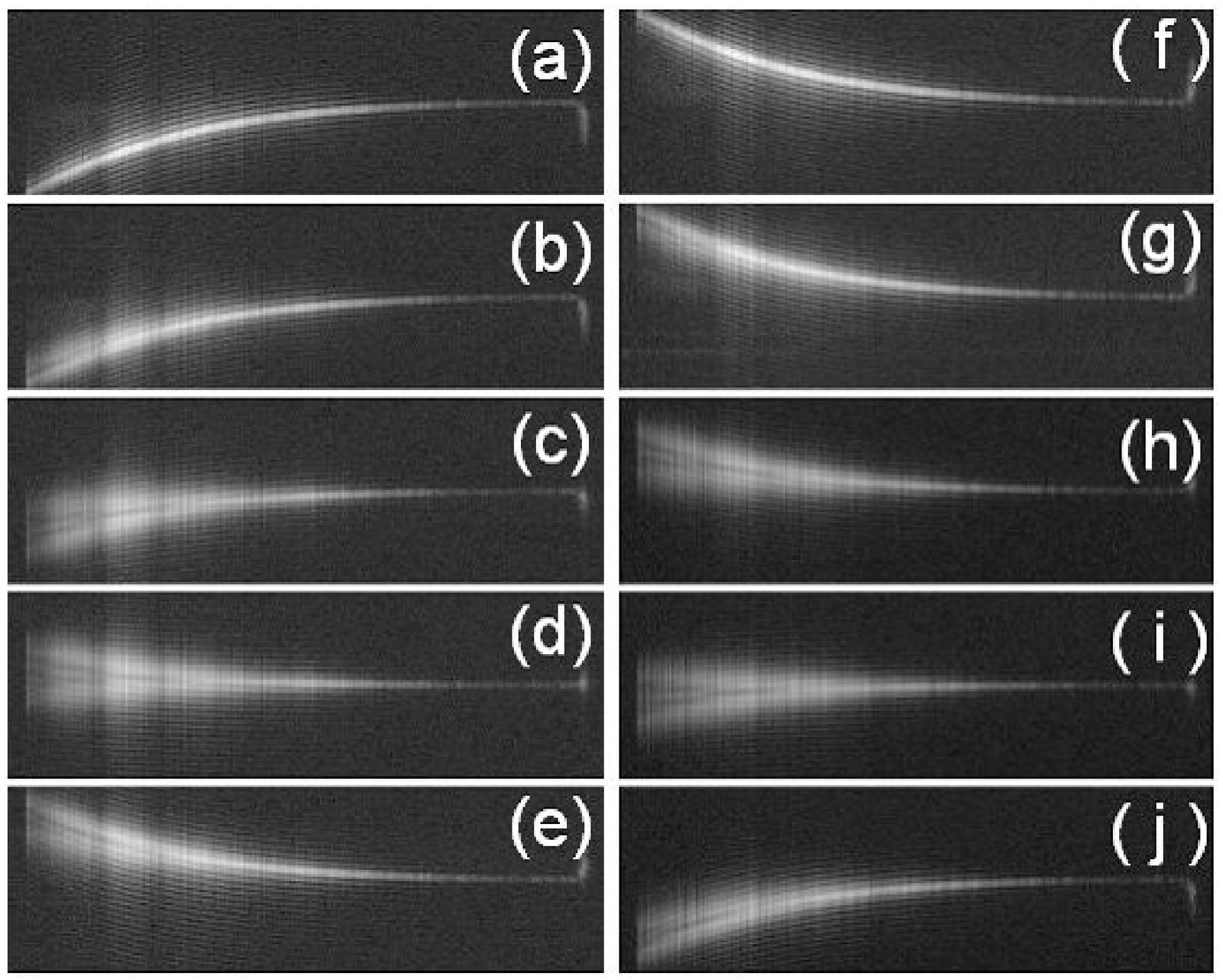}\\
  \caption{Successive positions of the reed on  a period $T_A$.  These images are obtained  by taking the section  in the $x = 268$ (horizontal white dashed line of Fig. \ref{Fig_images_log_xy_strobo} (e) )  of the stack of reconstructed images for $n =  -100$ to $ +100$. The images are displayed in logarithmic scale for the optical  field intensity $|E|^2$.}\label{Fig_images_xz_strobo}
\end{center}
\end{figure}

On each image, we get  signal   on the  zone $(x,y)$ of the reed where the instantaneous Doppler shift $2 V(x,y, x T_A )/\lambda$  is close to $n \nu_A$. This point is illustrated by Fig. \ref{Fig_velocity_field} that show as a function of the longitudinal axis $y$ the reed displacement $z(y,xT_A)$ and the reed velocity $V(y,xT_A)$ at illumination time $t=x T_A$.
The size, shape and brightness of the zone with signal depend    on the time of illumination $x T_A$ (i),  on the illumination  duration $\delta x T_A$ (ii), and  on the harmonic rank $n$ (iii).

For delay time $x T_A=0$, we observe  signal on a narrow bright zone of the reed for $n = 60$  (Fig. \ref{Fig_images_log_xy_strobo}(c) ), no signal for $n=-60$ (Fig. \ref{Fig_images_log_xy_strobo}(a) ) and some  signal for $n=0$ (Fig. \ref{Fig_images_log_xy_strobo}(b) ).  This means that  the reed is illuminated when it moves towards the detector, and  only the $n>0$ harmonic are generated, and detected. The narrow bright zone of Fig. \ref{Fig_images_log_xy_strobo}(c)  corresponds to the points of the reed where the instantaneous velocity at time $x T_A= 0$ is close to 
\begin{eqnarray}\label{Eq_77}
 V_{n=60}= 60 \nu_A c/ \nu_0
\simeq 8 \times 10^{-2}\textrm{ m}.\textrm{s}^{-1}
\end{eqnarray}
For time $ 0.3 T_A $, we observe a uniform signal on the reed for  $n = 0$ (Fig. \ref{Fig_images_log_xy_strobo}(e)) and no signal for $n=\pm 60$ (Fig. \ref{Fig_images_log_xy_strobo} (d,f) ). This  means that the reed is  in a position of maximal amplitude of  oscillation, with a velocity near zero.

For time $ 0.7 T_A$, the reed goes away from the detector. The results are similar to the ones obtained at time $x T_A=0$. We observe  a narrow bright zone for $n = - 60$ (Fig. \ref{Fig_images_log_xy_strobo}(g) ), no signal for $n=+60$ (Fig. \ref{Fig_images_log_xy_strobo} (i) ), and some  signal for $n=0$ (Fig. \ref{Fig_images_log_xy_strobo} (h) ).

To image  the reed instantaneous velocities during  vibration motion, we have swept the illumination time $x T_A$ from $0$ to $T_A$ by step of $0.1~T_A$. For each time $x T_A$, we have recorded the hologram for $n= -100$ to +100, and  we have stored the   $201$ reconstructed images in a 3D cube of data with axes $x$, $y$ and $n$.   The $n$ axis corresponds  to the Doppler shift, i.e. to the reed velocity.  We have performed  cuts of the 3D data in order to extract 2D images along axis $x$ and $n$  ($y$ being fixed). Figure \ref{Fig_images_xz_strobo}  displays the cuts obtained for $y=256$ , which correspond to the horizontal white dashed line in Fig.  \ref{Fig_images_log_xy_strobo} (e) ).

We actually obtain on Fig.  \ref{Fig_images_xz_strobo} a direct visualization of the shape of the reed instantaneous velocity, which varies with the illumination time $x T_A$. Since the motion is a sine function of time, the images of the instantaneous velocities of Fig.  \ref{Fig_images_xz_strobo} are similar to images of the reed itself, shifted in phase by $\pi/2$.
A movie, made  with the 10 images of Fig.  \ref{Fig_images_xz_strobo}, is provided in supplementary material (Media1.avi). Another movie, with 20 images per period, is also provided (Media2.avi). The movie shows the evolution of the reed velocities,  or the reed motion (if one neglects the shift of phase).

 %
%
%
%
%
%
%
%
%

We must notice that the images of  Fig.  \ref{Fig_images_xz_strobo}  (or the supplementary material movie) correspond to a huge amount of data, since it is necessary to record, for every  time $x T_A$, 4 images by harmonic rank $n$ with $n= -100$ to +100. We record thus $10 \times  4 \times 201 = 8040$ images. The frequency of acquisition of the camera being of 12.5 Hz,  the total recording time is approximately 12 minutes. It is necessary to add 
the   time  necessary for the calculation of reconstruction of holograms as well as the time needed  to  control, during acquisition, the change of  frequency of the synthesized signal generators SG2 and SG3. To get the 10 images of Fig.  \ref{Fig_images_xz_strobo}, the total time is thus about one hour.

\section{Conclusion}

This experiment  demonstrates that it is possible to  reconstruct a map of the instantaneous velocities  of a vibrating object by combining  sideband holography and stroboscopic   illumination synchronized with the vibration motion. Although the amount of generated data is huge, its acquisition is quite simple, since it is  fully automatized by using a computer that drives, through signal generators (SGs) and acousto optics modulators (AOMs), both the stroboscopic illumination, and the tuning of the sideband detection. The computer then performs both  data acquisition and image reconstruction.

The technique  is demonstrated here in the case of a simple sinusoidal oscillation. It can be extended to more complex periodic motions. One must notice that the technique is sensitive  to the direction of the instantaneous velocity (sign of $n$). The technique can thus be used to get the geometrical shape of a vibration mode in order to remove any ambiguity.


\begin{thebibliography}{}

\end{thebibliography}


\begin{thebibliography}{10}

\bibitem{pedrini1995digital}
G.~Pedrini, Y.L.~Zou, and H.J.~Tiziani,
\newblock "Digital double-pulsed holographic interferometry for vibration
  analysis,"
\newblock { Jour. of Mod. Opt.} \textbf{42}(2), 367--374 (1995)

\bibitem{pedrini1997digital}
G.~Pedrini, H.J. Tiziani, and Y.~Zou,
\newblock "Digital double pulse-tv-holography,"
\newblock {Opt. Laser Eng.} \textbf{26}(2),199--219 (1997)

\bibitem{pedrini1998transient}
G.~Pedrini, P.~Froening, H.~Fessler, and H.J.~Tiziani,
\newblock "Transient vibration measurements using multi-pulse digital
  holography,"
\newblock { Opt. Laser Tech.} \textbf{29}(8), 505--511 (1998)

\bibitem{pedrini2006high}
G.~Pedrini, W.~Osten, and M.E. Gusev,
\newblock "High-speed digital holographic interferometry for vibration
  measurement,"
\newblock { Appl. Opt. } \textbf{45}(15), 3456--3462 (2006)

\bibitem{fu2007vibration}
Y.~Fu, G.~Pedrini, and W.~Osten,
\newblock "Vibration measurement by temporal fourier analyses of a digital
  hologram sequence,"
\newblock { Appl. Opt. } \textbf{46}(23), 5719--5727 (2007)

\bibitem{powell1965iva}
R.L. Powell and K.A. Stetson,
\newblock {"Interferometric vibration analysis by wavefront reconstruction",}
\newblock { J. Opt. Soc. Am. } \textbf{55}(12), 1593--1598 (1965)

\bibitem{picart2003tad}
P.~Picart, J.~Leval, D.~Mounier, and S.~Gougeon,
\newblock {"Time-averaged digital holography,"}
\newblock {Opt. Lett.} \textbf{28}(20), 1900--1902 (2003)

\bibitem{pinard2003musical}
F.~Pinard, B.~Laine, and H.~Vach,
\newblock "Musical quality assessment of clarinet reeds using optical
  holography,"
\newblock { J. Acoust. Soc. Am.} \textbf{113}, 1736--1742   (2003)

\bibitem{zhang2004vap}
F.~Zhang, J.D.R. Valera, I.~Yamaguchi, M.~Yokota, and G.~Mills,
\newblock {"Vibration analysis by phase shifting digital holography,"}
\newblock { Opt. Rev.} \textbf{11}(5), 297--299 (2004)

\bibitem{demoli2004detection}
N.~Demoli and D.~Vukicevic,
\newblock "Detection of hidden stationary deformations of vibrating surfaces by
  use of time-averaged digital holographic interferometry,"
\newblock { Opt. Lett.} \textbf{29}(20), 2423--2425 (2004)

\bibitem{picart20052d}
P.~Picart, J.~Leval, J.C. Pascal, J.P. Boileau, M.~Grill, J.M. Breteau,
  B.~Gautier, and S.~Gillet,
\newblock "2d full field vibration analysis with multiplexed digital holograms,"
\newblock { Opt. Express} \textbf{13}(22), 8882--8892  (2005)

\bibitem{demoli2005dynamic}
N.~Demoli and I.~Demoli,
\newblock "Dynamic modal characterization of musical instruments using digital
  holography,"
\newblock { Opt. Express} \textbf{13}(13), 4812--4817  (2005)

\bibitem{demoli2006real}
N.~Demoli,
\newblock "Real-time monitoring of vibration fringe patterns by optical
  reconstruction of digital holograms: mode beating detection,"
\newblock { Opt. Express} \textbf{14}(6), 2117--2122 (2006)

\bibitem{asundi2006time}
A.~Asundi, V.R. Singh, 
\newblock "Time-averaged in-line digital holographic interferometry for
  vibration analysis,"
\newblock { Appl. Opt.} \textbf{45}(11), 2391--2395 (2006)

\bibitem{singh2007dynamic}
V.R. Singh, J.~Miao, Z.~Wang, G.~Hegde, and A.~Asundi,
\newblock "Dynamic characterization of mems diaphragm using time averaged
  in-line digital holography,"
\newblock { Opt. Comm.} \textbf{280}(2), 285--290 (2007)

\bibitem{picart2007tracking}
P.~Picart, J.~Leval, F.~Piquet, J.P. Boileau, T.~Guimezanes, and J.P. Dalmont,
\newblock "Tracking high amplitude auto-oscillations with digital fresnel
  holograms,"
\newblock { Opt. Express} \textbf{15}(13), 8263--8274 (2007)

\bibitem{fu2009vibration}
Y.~Fu, H.~Shi, and H.~Miao,
\newblock "Vibration measurement of a miniature component by high-speed
  image-plane digital holographic microscopy,"
\newblock { Appl. Opt.} \textbf{48}(11), 1990--1997 (2009)

\bibitem{kumar2009time}
U.P. Kumar, Y.~Kalyani, N.K. Mohan, and M.P. Kothiyal,
\newblock "Time-average tv holography for vibration fringe analysis,"
\newblock { Appl. Opt.} \textbf{48}(16), 3094--3101 (2009)

\bibitem{picart2005some}
P.~Picart, J.~Leval, D.~Mounier, and S.~Gougeon,
\newblock "Some opportunities for vibration analysis with time averaging in
  digital fresnel holography,"
\newblock { Appl. Opt.} \textbf{44}(3), 337--343 (2005)

\bibitem{borza2004high}
D.N. Borza,
\newblock "High-resolution time-average electronic holography for vibration
  measurement,"
\newblock {Opt. lasers Eng.} \textbf{41}(3), 515--527 (2004)

\bibitem{borza2005mechanical}
D.N. Borza,
\newblock "Mechanical vibration measurement by high-resolution time-averaged
  digital holography,"
\newblock { Meas. Sci. Technol.} \textbf{16}: 1853 (2005)

\bibitem{borza2006full}
D.N. Borza,
\newblock "Full-field vibration amplitude recovery from high-resolution
  time-averaged speckle interferograms and digital holograms by regional
  inverting of the bessel function,"
\newblock { Opt.  lasers  Eng.} \textbf{44}(8), 747--770  (2006)

\bibitem{joud2008imaging}
F.~Joud, F.~Lalo{\"e}, M.~Atlan, J.~Hare, and M.~Gross,
\newblock "Imaging of a vibrating object by sideband digital holography,"
\newblock { Opt. Express} \textbf{17}(4), 2774--2779  (2009)



\bibitem{aleksoff1971temporally}
CC~Aleksoff,
\newblock "Temporally modulated holography,"
\newblock { Appl. Opt.} \textbf{10}(6), 1329--1341  (1971)

\bibitem{le2000numerical}
F.~Le~Clerc, L.~Collot, and M.~Gross,
\newblock "Numerical heterodyne holography with two-dimensional photodetector
  arrays,"
\newblock { Opt. Lett.} \textbf{25}(10), 716--718 (2000)

\bibitem{gross2003shot}
M.~Gross, P.~Goy, and M.~Al-Koussa,
\newblock "Shot-noise detection of ultrasound-tagged photons in
  ultrasound-modulated optical imagin,"
\newblock { Opt. Lett.} \textbf{28}(24), 2482--2484 (2003)

\bibitem{joud2009fringe}
F.~Joud, F.~Verpillat, F.~Lalo{\"e}, M.~Atlan, J.~Hare, and M.~Gross,
\newblock "Fringe-free holographic measurements of large-amplitude vibrations,"
\newblock { Opt. Lett.} \textbf{34}(23), 3698--3700 (2009)

\bibitem{leval2005full}
J.~Leval, P.~Picart, J.P. Boileau, and J.C. Pascal,
\newblock "Full-field vibrometry with digital fresnel holography,"
\newblock { Appl. Opt.} \textbf{44}(27), 5763--5772 (2005)

\bibitem{schnars1994direct}
U.~Schnars and W.~J{\"u}ptner,
\newblock "Direct recording of holograms by a ccd target and numerical
  reconstruction,"
\newblock { Appl. Opt.} \textbf{33}(2), 179--181  (1994)

\bibitem{picart2008general}
P.~Picart and J.~Leval,
\newblock General theoretical formulation of image formation in digital fresnel
  holography.
\newblock { J. Opt. Soc. Am. A} \textbf{25}(7), 1744--1761 (2008)

\end{thebibliography}
\end{document}